\documentclass[12pt]{iopart}
\usepackage{iopams}  
\usepackage{setstack}

\begin{document}
\title[Discrete exponential type systems on a quad graph]{Discrete exponential type systems on a quad graph, corresponding to the affine Lie algebras $A^{(1)}_{N-1}$.}

\author{I T Habibullin$^{1,2}$, A R Khakimova$^{1}$}

\address{$^1$Institute of Mathematics, Ufa Federal Research Centre, Russian Academy of Sciences,
112, Chernyshevsky Street, Ufa 450008, Russian Federation}
\address{$^2$Bashkir State University, 32, Validy Street, Ufa 450076 , Russian Federation} 
\eads{\mailto{habibullinismagil@gmail.com}, \mailto{aigul.khakimova@mail.ru}}

\begin{abstract}
The article deals with the problem of the integrable discretization of the well-known Drinfeld-Sokolov hierarchies related to the Kac-Moody algebras. A class of discrete exponential systems connected with the Cartan matrices has been suggested earlier in \cite{GHY} which coincide with the corresponding Drinfeld-Sokolov systems in the continuum limit. It was conjectured that the systems in this class are all integrable and the conjecture has been approved by numerous examples. In the present article we study those systems from this class which are related to the algebras $A^{(1)}_{N-1}$. We found the Lax pair for arbitrary $N$, briefly discussed the possibility of using the method of formal diagonalization of Lax operators for describing a series of local conservation laws and illustrated the technique using the example of $N=3$. Higher symmetries of the system $A^{(1)}_{N-1}$ are presented in both characteristic directions. Found recursion operator for $N=3$. It is interesting to note that this operator is not weakly nonlocal.

\end{abstract}


\maketitle

\eqnobysec

\section{Introduction}
Exponential type systems of hyperbolic equations in partial derivatives
\begin{equation}\label{hypPDE} 
v^{i}_{x,y}=\exp({a_{i1}v^{1}+a_{i2}v^{2}+\cdots +a_{iN}v^{N}}),\, 1\leq i\leq N
\end{equation}
are actively discussed in literature due to their applications in the field theory and the other areas of physics, as well as in geometry, the integrability theory etc. 

Exponential system in the form of the two dimensional Toda lattice 
\begin{equation}\label{TL} 
v^{i}_{x,y}=\exp({v^{i+1}-2v^{i}+v^{i-1}}) 
\end{equation} 
has appeared many years ago within the frame of the Laplace cascade integration method (see \cite{Darboux}). In the context of the soliton theory this equation has been rediscovered in \cite{Toda}, \cite{Mikhailov}. Due to the works by Mikhailov, Olshanetsky, Perelomov, Leznov, Savel'ev, Shabat, Smirnov, Wilson, Yamilov, Drinfeld, Sokolov and many others mathematical theory of the exponential systems has been developed and nowadays (see \cite{Mikhailov}-\cite{Kac})  it is well-known that in the case when the coefficient matrix $A=\{a_{i,j}\}$ coincides with the Cartan matrix of a semi-simple Lie algebra then (\ref{hypPDE}) admits a complete set of non-trivial integrals in both characteristic directions and therefore is integrable in the sense of Darboux. Similarly, if $A$ is the generalized Cartan matrix of an affine Lie algebra then the system  (\ref{hypPDE}) can be studied by means of the inverse scattering transform method \cite{Mikhailov}-\cite{Kac}. 

Inspired by Ward (see \cite{Ward}) the problem of finding integrable discrete analogs of equation (\ref{hypPDE}), which are  discrete in both independent variables,  has been intensively studying since the middle of 90's. We remark that the particular $1+1$ dimensional case, obtained by setting $x=y$ was successfully investigated in \cite{Suris}.
An effective way to construct the discrete version of (\ref{TL}) based on the discretization of the bilinear equation of the Toda lattice, was suggested by  Hirota \cite{Hirota} and Miwa \cite{Miwa}. Hirota-Miwa equation is a universal soliton equation from which a large variety of integrable discrete models can be derived by symmetry constraints (see \cite{Willox}). An alternative approach to discretise the two dimensional Toda lattice is used by  Fordy and Gibbons \cite{Gibbons80}, \cite{Gibbons83}, where the discrete equation is derived as the superposition formula of the Backlund transformations for the equation (\ref{TL}). A large class of the quad systems and their applications in physics are studied in \cite{Kuniba}.

In the article \cite{GHY} a class of exponential type systems of discrete equations 
\begin{equation}\label{generalghy}
ae^{-u^i_{1,1}+u^i_{0,1}+u^i_{1,0}-u^i_{0,0}}-1=b\exp\left(\sum_{j=1}^{i-1}a_{i,j}u^j_{1,0}+\sum_{j=i+1}^{N}a_{i,j}u^j_{0,1}+a_{i,i}\frac{u^i_{0,1}+u^i_{1,0}}{2}\right)
\end{equation}
has been suggested for the particular case $a=b=1$. In the present article we assume that $a$ and $b$ are arbitrary nonzero constants. Here the upper index $j$ is a natural ranging from $1$ to $N$. For the shifts of the sought functions $u^j_{n,m}$ in (\ref{generalghy}) we  use the abbreviated notations $u^j_{n+i,m+k}=u^j_{i,k}$ such that $u^j_{1,0}$ means $u^j_{n+1,m}$ and so on. 

For the small values of $N$  system (\ref{generalghy}) was obtained in \cite{GHY} from integrable cases of (\ref{hypPDE}) by applying the method of discretization preserving integrals and symmetries. For (\ref{generalghy}) with $A$ being the generalized Cartan matrix of the algebra $D^{(2)}_N$ the Lax pair has been found. Later in  \cite{Smirnov} S.V.Smirnov proved that the quad systems (\ref{generalghy}) corresponding to simple Lie algebras $A_N$ and $B_N$ are Darboux integrable. These facts partially confirm our hypothesis from \cite{GHY} that the quad system inherits the integrability property of the system (\ref{hypPDE}) and is integrable for the Cartan matrices of both simple and affine Lie algebras.

The main goal of this paper is to present the Lax pair for the quad system (\ref{generalghy}) associated with the algebra $A^{(1)}_{N-1}$. We outlined a method for applying the formal diagonalization technique for finding local conservation laws and higher symmetries. In other words we approve the aforementioned hypothesis for one more series of affine Lie algebras. Unfortunately, our Lax pairs for $A^{(1)}_{N-1}$ and $D^{(2)}_N$ are not given in terms of the Cartan–-Weyl basis of the algebra as it was in the case of  (\ref{hypPDE}) (see \cite{Drinfeld}),  so there are problems with generalization to other algebras.
Note that alternative examples of discretizations of the Toda lattices related with the algebra $A^{(1)}_{N-1}$ are investigated in \cite{Xenitidis}-\cite{Habibullin}.

Quadrilateral lattices studied in the article are very close to those suggested in \cite{Nijhoff} as discretizations of the Gelfand-Dikii hierarchy. However these two classes of the discrete equations  differ from each other. For instance, the first members are the discrete d'Alembert equation and, respectively discrete potential KdV equation. The second members are the couplet system (\ref{eqA11}) and the discrete Boussinesq equation.  They are not connected by the point transformations. 

An interesting integrable system of partial difference equations is suggested in \cite{Doliwa} with the arbitrarily large number of the independent variables.  In a particular case when the number of independent variables is two this system looks very similar to (\ref{eqA1N}) and (\ref{periodic_system_r}), but the systems are not equivalent (see Proposition 3 in \S4).

Let's briefly discuss the content of the article. In \S2 we study the quasi-periodic reduction $t^{j+N}_{n,m}=t^j_{n+1,m-1}$ of the Hirota-Miwa equation which leads to a quadrilateral system of the form (\ref{generalghy}) corresponding to the algebra $A^{(1)}_{N-1}$. We show that the constraint $\psi^{j+N}_{n,m}=\lambda\psi^j_{n+1,m-1}$ imposed on the system (\ref{lax1}) associated with the Hirota-Miwa equation successfully creates a Lax pair for the resulting  quad system written in the form (\ref{A1-N}). 

We also give an example of a system obtained using the purely periodic condition $t^{j+N}_{n,m}=t^j_{n,m}$. Note that various types of the periodic conditions for the Hirota-Miwa equation are studied in \cite{Zabrodin}. For example, it is shown that the restriction $t^{j+2}_{n,m}=t^j_{n,m}$ leads to a discrete version of the sine-Gordon equation. However, to our knowledge, examples of quasi-periodic reduction of the form $t^{j+N}_{n,m}=t^j_{n+1,m-1}$ are not considered earlier.

In \S3 we used the Lax pair derived in the previous section for describing the local conservation laws for the system (\ref{A1-N}). To this end we converted the Lax equations (\ref{7new}), (\ref{8new}) around each singular value of the spectral parameter $\lambda$ to a special form allowing to determine asymptotic representation of the eigenfunctions. Finding of this special form usually causes the main problem. In section 3 we found triangular transformations reducing the Lax equations to the required form. The case $N=3$ is studied in more details. We note that for constructing the asymptotic representation of the Lax eigenfunctions we used a method which slightly differs from known ones (see \cite{HabYang}, \cite{Mikhailov15}).

In the forth section  higher symmetries are given for the quad system (\ref{A1-N}), and finally in \S5 the recursion operators are found from the asymptotic representations of the Lax eigenfunctions for $N=2$ and $N=3$ in the particular values of the parameters $a=b=1$. For the case $N=3$ the found recursion operator is not weakly nonlocal.

\section{Derivation of the Lax pair}

\subsection{Quasi-periodic reduction}

In the article we deal with the quad system (\ref{generalghy}) corresponding to the algebra $A^{(1)}_{N-1}$ which can be written in terms of the variable $t^j_{n,m}=\exp\left\{-u^j_{n,m}\right\}$ as follows 
\begin{eqnarray}\label{A1-N}
at^1_{0,0}t^1_{1,1}-t^1_{1,0}t^1_{0,1}=bt^{N}_{0,1}t^{2}_{0,1},\nonumber\\
at^j_{0,0}t^j_{1,1}-t^j_{1,0}t^j_{0,1}=bt^{j-1}_{1,0}t^{j+1}_{0,1}, \quad 2\leq j\leq N-1,\\
at^N_{0,0}t^N_{1,1}-t^N_{1,0}t^N_{0,1}=bt^{N-1}_{1,0}t^{1}_{1,0}.\nonumber
\end{eqnarray}
Evidently system (\ref{A1-N}) can be obtained from the Hirota-Miwa equation 
\begin{eqnarray}\label{HM}
at^j_{0,0}t^j_{1,1}-t^j_{1,0}t^j_{0,1}=bt^{j-1}_{1,0}t^{j+1}_{0,1}, \quad -\infty\leq j\leq +\infty
\end{eqnarray}
by imposing the quasi-periodicity closure constraint
\begin{eqnarray}\label{quasi_original}
t^{j+N}_{0,0}=t^j_{1,-1}, \quad N\geq 2.
\end{eqnarray} 
Note that for the simplest case $N=1$ (\ref{HM}) implies the discrete version of the d'Alembert equation.

Our aim is to derive the Lax pair for (\ref{A1-N}) from the overdetermined system of the linear equations
\begin{equation}\label{lax1}
\psi^j_{1,0}=\frac{t^{j+1}_{1,0}t^j_{0,0}}{t^{j+1}_{0,0}t^j_{1,0}}\psi^j_{0,0}-\psi^{j+1}_{0,0}, \quad
\psi^j_{0,1}=\psi^{j}_{0,0}+b\frac{t^{j+1}_{0,1}t^{j-1}_{0,0}}{t^j_{0,0}t^j_{0,1}}\psi^{j-1}_{0,0}
\end{equation}
associated with the equation (\ref{HM}). More precisely, when $t^j_{n,m}$ solves equation (\ref{A1-N}) then the system (\ref{lax1}) is compatible. However the converse is not true: the compatibility of the system does not imply (\ref{HM}). Nevertheless (\ref{lax1}) can be used effectively for constructing the true Lax pair for the quad system (\ref{A1-N}). Evidently (\ref{lax1}) implies the hyperbolic type discrete linear equation
\begin{equation}\label{eqn1.1}
\psi^j_{1,1}-\psi^j_{1,0}-\frac{t^{j+1}_{1,1}t^j_{0,1}}{t^{j+1}_{0,1}t^j_{1,1}}\psi^j_{0,1}+
a\frac{t^j_{0,0}t^{j+1}_{1,1}}{t^j_{1,0}t^{j+1}_{0,1}}\psi^j_{0,0}=0.
\end{equation}
It is widely known that the Laplace invariants are important characteristics  of the hyperbolic type discrete and continuous equations. Recall that for an equation of the form 
\begin{equation}\label{f}
f_{1,1}+b_{0,0}f_{1,0}+c_{0,0}f_{0,1}+d_{0,0}f_{0,0}=0
\end{equation} 
the Laplace invariants $K_1$ and $K_2$ are determined due to the rules (see \cite{NovikovDynnikov,AdlerStartsev})
\begin{equation*}\label{LaplaceInvar}
K_1=\frac{b_{0,0}c_{1,0}}{d_{1,0}},\quad 
K_2=\frac{b_{0,1}c_{0,0}}{d_{0,1}}.
\end{equation*}
Further we will use theorem (see \cite{AdlerStartsev}) claiming that equation (\ref{f}) and the equation
\begin{equation*}\label{tildef}
\tilde f_{1,1}+\tilde b_{0,0} \tilde f_{1,0}+\tilde c_{0,0}\tilde f_{0,1}+\tilde d_{0,0}\tilde f_{0,0}=0
\end{equation*} 
are related with one another by the multiplicative transformation $f=\lambda\tilde f$ if and only if they have the same pair of the Laplace invariants, i.e. $K_1=\tilde K_1$ and $K_2=\tilde K_2$.

{\bf Proposition 1.} Under the quasi-periodicity condition (\ref{quasi_original}) equation (\ref{eqn1.1}) and the equation
\begin{equation}\label{eqn1.1N}
\psi^{j+N}_{1,1}-\psi^{j+N}_{1,0}-\frac{t^{j+N+1}_{1,1}t^{j+N}_{0,1}}{t^{j+N+1}_{0,1}t^{j+N}_{1,1}}\psi^{j+N}_{0,1}+
a\frac{t^{j+N}_{0,0}t^{j+N+1}_{1,1}}{t^{j+N}_{1,0}t^{j+N+1}_{0,1}}\psi^{j+N}_{0,0}=0
\end{equation}
are related by the multiplicative transformation, or more precisely, by the equation
\begin{equation}\label{multiplicative}
\psi^{j}_{1,0}=A^j\psi^{j+N}_{0,1}.
\end{equation}

{\bf Proof}. Let us first give the Laplace invariants in enlarged (not abbreviated!) form
$$K_{1}(n,m,j)=\frac{t^j_{n+2,m}t^j_{n+1,m+1}}{t^j_{n+1,m}t^j_{n+2,m+1}a},\quad K_{2}(n,m,j)=\frac{t^{j+1}_{n+1,m+1}t^{j+1}_{n,m+2}}{t^{j+1}_{n,m+1}t^{j+1}_{n+1,m+2}a}.\quad\\
$$
Now it is easily seen that constraint (\ref{quasi_original}) implies $K_1(n+1,m,j)=K_1(n,m+1,j+N)$ and $K_2(n+1,m,j)=K_2(n,m+1,j+N)$. These two relations due to the over-mentioned theorem allow one to complete the proof.

{\bf Proposition 2}. The coefficient $A^j$ in the relation (\ref{multiplicative}) does not depend on any of the variables $j,n,m$.

{\bf Proof}. By applying the shift operator $D_m$, acting according to the rule $D_my_m=y_{m+1}$ to equation (\ref{multiplicative}), we evidently obtain $\psi^{j}_{1,1}=A^j_{0,1}\psi^{j+N}_{0,2}$. Next we replace $\psi^j_{1,1}$ due to the hyperbolic type equation (\ref{eqn1.1}) and find
$$\psi^j_{1,0}+\frac{t^{j+1}_{1,1}t^j_{0,1}}{t^{j+1}_{0,1}t^j_{1,1}}\psi^j_{0,1}-a\frac{t^j_{0,0}t^{j+1}_{1,1}}{t^j_{1,0}t^{j+1}_{0,1}}\psi^j_{0,0}=A^j_{0,1}\psi^{j+N}_{0,2}.$$
Then we get rid the function $\psi^j$ by virtue of the relation (\ref{multiplicative}). As a result we arrive at the equation
\begin{equation*}\label{eqn1.1NA}
\psi^{j+N}_{1,1}-\frac{A^j_{1,-2}}{A^j_{1,0}}\psi^{j+N}_{1,0}-\frac{A^j_{0,0}}{A^j_{1,0}}\frac{t^{j+1}_{2,0}t^{j}_{1,0}}{t^{j+1}_{1,0}t^{j}_{2,0}}\psi^{j+N}_{0,1}+a\frac{A^j_{0,-1}}{A^j_{1,0}}\frac{t^{j}_{1,-1}t^{j+1}_{2,0}}{t^{j}_{2,-1}t^{j+1}_{1,0}}\psi^{j+N}_{0,0}=0
\end{equation*}
which should coincide with (\ref{eqn1.1N}). Now we compare the corresponding coefficients of these two equations and by means of (\ref{multiplicative}) we get three equations for $A^j$
$$A^j_{1,-2}=A^j_{1,0}, \quad A^j_{1,0}=A^j_{0,0},\quad A^j_{0,-1}=A^j_{1,0}$$
which approve that $A^j$ does not depend on $n$ and $m$.

It obviously follows from (\ref{multiplicative}) that

\begin{equation}\label{Aj}
\psi^j_{1,0}-\frac{t^{j+1}_{1,0}t^j_{0,0}}{t^{j+1}_{0,0}t^j_{1,0}}\psi^j_{0,0}=A^j\left( \psi^{j+N}_{0,1}-\frac{t^{j+1}_{1,0}t^j_{0,0}}{t^{j+1}_{0,0}t^j_{1,0}}\psi^{j+N}_{-1,1}  \right). 
\end{equation}
The left hand side of (\ref{Aj}) coincides with $-\psi^{j+1}_{0,0}$. Let us replace the fraction  by means of the relation 
(\ref{quasi_original}) and find that due to (\ref{lax1}) the right hand side in (\ref{Aj}) coincides with $-A^j\psi^{j+N+1}_{-1,1}$. Thus we have a relation $\psi^{j+1}_{1,0}=A^j\psi^{j+N+1}_{0,1}$ which together with (\ref{multiplicative}) gives $A^j=A^{j+1}$. Proposition 2 is proved.

Let us consider quad system (\ref{A1-N}). It is easily verified that (\ref{A1-N}) can be quasi-periodically prolonged to the infinite interval $-\infty<j<+\infty$ by setting $t^j_{1,0}=t^{j+N}_{0,1}$ such that the prolonged function $t^j_{n,m}$ will solve equation (\ref{HM}). Then the Propositions 1,2 provide the following gluing conditions
\begin{equation*}\label{glue}
\psi^0_{1,0}=\lambda^{-1} \psi^{N}_{0,1}\quad \mbox{and}\quad \psi^{N+1}_{0,1}=\lambda\psi^1_{1,0},
\end{equation*}
where $\lambda$ is an arbitrary parameter.

In order to derive the Lax pair for the quad system (\ref{A1-N}) we impose the gluing conditions on the linear system (\ref{lax1}) and find 
\begin{eqnarray}\label{7}
\psi^j_{1,0}=\frac{t_{1,0}^{j+1}t^j_{0,0}}{t^{j+1}_{0,0}t^j_{1,0}}\psi^j_{0,0}-\psi^{j+1}_{0,0}, \, 1\leq j\leq N-1,\\
\psi^N_{1,0}=\frac{t_{1,0}^{N+1}t^N_{0,0}}{t^{N+1}_{0,0}t^N_{1,0}}\psi^N_{0,0}-\lambda\psi^{1}_{1,-1}\nonumber
\end{eqnarray}
and
\begin{eqnarray}\label{8}
\psi^1_{0,1}=\psi^1_{0,0}+b\lambda^{-1}\frac{t^2_{0,1}t^0_{0,0}}{t^{1}_{0,1}t^1_{0,0}}\psi^{N}_{-1,1},\\
\psi^j_{0,1}=\psi^j_{0,0}+b\frac{t^{j+1}_{0,1}t^{j-1}_{0,0}}{t^{j}_{0,1}t^j_{0,0}}\psi^{j-1}_{0,0}, \quad 2\leq j\leq N. \nonumber
\end{eqnarray}

Let us apply now the shift operator $D_m$ to the last equation in (\ref{7}):
\begin{equation*}
\psi^N_{1,1}=\frac{t_{1,1}^{N+1}t^N_{0,1}}{t^{N+1}_{0,1}t^N_{1,1}}\psi^N_{0,1}-\lambda\psi^{1}_{1,0}.
\end{equation*}
Due to the equation (\ref{eqn1.1}) taken at the value $j=N$ we obtain after substitution $t^{N+1}_{1,1}=t^1_{2,0}$, $t^{N+1}_{0,1}=t^1_{1,0}$ that
\begin{equation*}
\psi^N_{1,0}-a\frac{t^{N}t^1_{2,0}}{t^{N}_{1,0}t^1_{1,0}}\psi^N_{0,0}=-\lambda\psi^{1}_{1,0}.
\end{equation*}
By replacing $\psi^1_{1,0}$ from (\ref{7}) we bring the equation to the suitable form
\begin{equation*}
\psi^N_{1,0}=-\lambda\frac{t^{2}_{1,0}t^1_{0,0}}{t^2_{0,0} t^1_{1,0}}\psi^1_{0,0}+\lambda\psi^{2}_{0,0} + a\frac{t^N_{0,0}t^1_{2,0}}{t_{1,0}^Nt^1_{1,0}}\psi^N_{0,0}.  
\end{equation*}

In the same way we can rewrite in the appropriate form the first equation in (\ref{8}). Since the reasonings are very similar to that used above we omit them and give only the final result 
\begin{eqnarray*}
\psi^1_{0,1}=a\frac{t^1_{0,0}t^1_{1,1}}{t^1_{1,0}t^1_{0,1}}\psi^1_{0,0}+b^2\lambda^{-1}\frac{t^{N-1}_{0,0}t^2_{0,1}}{t^N_{0,0}t^1_{0,1}}\psi^{N-1}_{0,0}+b\lambda^{-1}\frac{t^2_{0,1}t^N_{0,1}}{t^1_{1,0}t^1_{0,1}}\psi^{N}_{0,0}.
\end{eqnarray*}

Let us summarize the computations above and present the desired Lax pair: 
\begin{eqnarray}\label{7new}
\psi^j_{1,0}=\frac{t_{1,0}^{j+1}t^j_{0,0}}{t^{j+1}_{0,0}t^j_{1,0}}\psi^j_{0,0}-\psi^{j+1}_{0,0}, \, 1\leq j\leq N-1,\\
\psi^N_{1,0}=-\lambda\frac{t^{2}_{1,0}t^1_{0,0}}{t^2_{0,0} t^1_{1,0}}\psi^1_{0,0}+\lambda\psi^{2}_{0,0} + a\frac{t^N_{0,0}t^1_{2,0}}{t_{1,0}^Nt^1_{1,0}}\psi^N_{0,0}\nonumber
\end{eqnarray}
and
\begin{eqnarray}\label{8new}
\psi^1_{0,1}=a\frac{t^1_{0,0}t^1_{1,1}}{t^1_{1,0}t^1_{0,1}}\psi^1_{0,0}+b^2\lambda^{-1}\frac{t^{N-1}_{0,0}t^2_{0,1}}{t^N_{0,0}t^1_{0,1}}\psi^{N-1}_{0,0}+b\lambda^{-1}\frac{t^2_{0,1}t^N_{0,1}}{t^1_{1,0}t^1_{0,1}}\psi^{N}_{0,0},\\
\psi^j_{0,1}=\psi^j_{0,0}+b\frac{t^{j+1}_{0,1}t^{j-1}_{0,0}}{t^{j}_{0,1}t^j_{0,0}}\psi^{j-1}_{0,0}, \quad 2\leq j\leq N. \nonumber
\end{eqnarray}

Surprisingly the Lax pair (\ref{7new}), (\ref{8new}) is obtained from (\ref{lax1}) by changing only two equations. 

\subsection{Periodic reduction}

Let's briefly discuss the periodic closure condition (see also \cite{Zabrodin})
\begin{equation}\label{periodic}
t^{j+N}_{0,0}=t^j_{0,0}
\end{equation}
for the Hirota-Miwa equation (\ref{HM}). It is easily checked that (\ref{periodic}) generates a closure constraint for $\psi$:
\begin{equation*}\label{periodic_psi}
\psi^{j+N}_{0,0}=\xi\psi^j_{0,0}.
\end{equation*}
As a result we get a quad system
\begin{eqnarray}\label{periodic_system}
at^1_{0,0}t^1_{1,1}-t^1_{1,0}t^1_{0,1}=bt^{N}_{1,0}t^{2}_{0,1},\nonumber\\
at^j_{0,0}t^j_{1,1}-t^j_{1,0}t^j_{0,1}=bt^{j-1}_{1,0}t^{j+1}_{0,1}, \quad 2\leq j\leq N-1,\\
at^N_{0,0}t^N_{1,1}-t^N_{1,0}t^N_{0,1}=bt^{N-1}_{1,0}t^{1}_{0,1}.\nonumber
\end{eqnarray}
In this case, the corresponding reduction of linear equations (\ref{lax1}) is found immediately
\begin{equation}\label{periodic_lax}
\Phi_{1,0}=F\Phi, \quad \Phi_{0,1}=G\Phi,
\end{equation}
where $\Phi=\left(\psi^1,\psi^2,\dots,\psi^N\right)^{T}$ and
\begin{eqnarray*}
F=\left(
\begin{array}{ccccc}
\frac{t^1_{0,0}t^2_{1,0}}{t^1_{1,0}t^2_{0,0}}&-1&0&\dots&0\\
0&\frac{t^2_{0,0}t^{3}_{1,0}}{t^2_{1,0}t^{3}_{0,0}}&-1&\dots&0\\
0&0&\frac{t^3_{0,0}t^{4}_{1,0}}{t^3_{1,0}t^{4}_{0,0}}&\dots&0\\
&&&\dots\\
-\xi&0&\dots&0&\frac{t^{N}_{0,0}t^1_{1,0}}{t^{N}_{1,0}t^1_{0,0}}\\
\end{array}\right),
\end{eqnarray*}
\begin{eqnarray*}
G=\left(
\begin{array}{ccccc}
1&0&\dots&0&\xi^{-1}\frac{bt^2_{0,1}t^N_{0,0}}{t^1_{0,0}t^1_{0,1}}\\
\frac{bt^1_{0,0}t^3_{0,1}}{t^2_{0,0}t^2_{0,1}}&1&\dots&0&0\\
0&\frac{bt^2_{0,0}t^4_{0,1}}{t^3_{0,0}t^3_{0,1}}&\dots&0&0\\
&&\dots\\
0&0&\dots&\frac{bt^{N-1}_{0,0}t^1_{0,1}}{t^{N}_{0,0}t^{N}_{0,1}}&1\\
\end{array}\right).
\end{eqnarray*}

It can be checked by a direct computation that system (\ref{periodic_lax}) does not define a Lax pair for (\ref{periodic_system}) (In contrast to the quasi-periodic case (\ref{quasi_original})). More precisely, the consistency of (\ref{periodic_lax}) does not imply (\ref{periodic_system}). However the situation is changed if we pass in (\ref{periodic_system}) to the potential variables $r^j_{0,0}=\frac{t^j_{1,0}}{t^j_{0,0}}$:
\begin{equation} \label{periodic_system_r}
\left. \begin{array}{l}
ar^1_{1,1}=r^1_{1,0}+r^2_{0,1}r^N_{1,0}\left(\frac{a}{r^1_{0,0}}-\frac{1}{r^1_{0,1}}\right),\\
ar^j_{1,1}=r^j_{1,0}+r^{j-1}_{1,0}r^{j+1}_{0,1}\left(\frac{a}{r^j_{0,0}}-\frac{1}{r^j_{0,1}}\right), \quad 2\leq j\leq N-1,\\
ar^N_{1,1}=r^N_{1,0}+r^{N-1}_{1,0}r^{1}_{0,1}\left(\frac{a}{r^N_{0,0}}-\frac{1}{r^N_{0,1}}\right).
\end{array} \right.
\end{equation}
Now the system (\ref{periodic_lax}) with the potentials
\begin{eqnarray*}
F=\left(
\begin{array}{ccccc}
\frac{r^2_{0,0}}{r^1_{0,0}}&-1&0&\dots&0\\
0&\frac{r^3_{0,0}}{r^2_{0,0}}&-1&\dots&0\\
0&0&\frac{r^4_{0,0}}{r^3_{0,0}}&\dots&0\\
&&&\dots\\
-\xi&0&\dots&0&\frac{r^1_{0,0}}{r^N_{0,0}}\\
\end{array}\right),
\end{eqnarray*}
\begin{eqnarray*}
G=\left(
\begin{array}{ccccc}
1&0&\dots&0&\xi^{-1}\frac{ar^1_{0,1}-r^1_{0,0}}{r^N_{0,0}}\\
\frac{ar^2_{0,1}-r^2_{0,0}}{r^1_{0,0}}&1&\dots&0&0\\
0&\frac{ar^3_{0,1}-r^3_{0,0}}{r^2_{0,0}}&\dots&0&0\\
&&\dots\\
0&0&\dots&\frac{ar^{N}_{0,1}-r^{N}_{0,0}}{r^{N-1}_{0,0}}&1\\
\end{array}\right)
\end{eqnarray*}
provides the Lax pair for (\ref{periodic_system_r}). 


\section{Formal asymptotic solutions of the direct scattering problem and the local conservation laws}

The method of the formal diagonalization of the Lax pairs given by differential operators is suggested in \cite{Drinfeld}. It is based on the ideas and technique applied earlier \cite{Wasow} in order to construct asymptotic solutions to the systems of differential equations with a parameter, when the parameter goes to its singular value. Let us recall that the formal diagonalization provides a main step in solving the direct scattering problem and allows us to describe local conservation laws and higher symmetries. 

For the system of the linear discrete equations with a parameter two different tools to construct formal asymptotics has been suggested in \cite{HabYang} and \cite{Mikhailov15}. However we do not see how to apply either of them to the case of the systems (\ref{7new}), (\ref{8new}) since the method from \cite{HabYang} needs a special form of the potential and that of \cite{Mikhailov15} can be applied only in the case when the corresponding quad equation admits evolutionary type higher symmetries. That is why we are forced to suggest an alternative way of the formal diagonalization which slightly differs from the above-mentioned ones. Below we briefly discuss the scheme.

Let us consider a system of the discrete linear equations
\begin{equation}  \label{lineardiscrete}
Y_{n+1}=f_nY_n, \quad f_n=\sum^{\infty}_{j=-1}f_n^{(j)}\lambda^{-j},
\end{equation}
where $f_n^{(j)}\in{\bf C}^{k\times k}$ for $j\geq-1$ are matrix valued functions.
 In order to identify the matrix structure of the potential we divide the matrices into blocks as
\begin{equation*} \label{A}
A=\left( \begin{array}{cc}
A_{11}&A_{12}\\
A_{21}&A_{22}
\end{array} \right),
\end{equation*}
where the blocks $A_{11}$, $A_{22}$ are square matrices. Here we assume that in (\ref{lineardiscrete}) the coefficient $f_n^{(-1)}$ is of one of the forms
\begin{equation} \label{f_01}
f_n^{(-1)}=\left( \begin{array}{cc}
0&0\\
0&A_{22}
\end{array} \right),\quad \det A_{22}\neq 0
\end{equation}
or 
\begin{equation} \label{f_02}
f_n^{(-1)}=\left( \begin{array}{cc}
A_{11}&0\\
0&0
\end{array} \right), \quad \det A_{11}\neq 0.
\end{equation}
Now our goal is to bring (\ref{lineardiscrete}) to a block-diagonal form 
\begin{equation}\label{diag}
\varphi_{n+1}=h_n\varphi_n,
\end{equation}
where $h_n$ is a formal series
\begin{equation}\label{series1}
h_n=h_n^{(-1)}\lambda +  h_n^{(0)}+h_n^{(1)}\lambda^{-1}+h_n^{(2)}\lambda^{-2}+\cdots
\end{equation}
with the coefficients having the block structure
\begin{equation} \label{coeff-h}
h_n^{(j)}=\left( \begin{array}{cc}
h^{(j)}_{11}&0\\
0&h^{(j)}_{22}
\end{array} \right).
\end{equation}
To this end we use the linear transformation $Y_n=T_n\varphi_n$ assuming that $T_n$ is also a formal series
\begin{equation*}
T_n=E+T_n^{(1)}\lambda^{-1}+T_n^{(2)}\lambda^{-2}+\cdots. \label{series2}
\end{equation*}
where $E$ is the unity matrix and $T_n^{(j)}$ is a matrix with vanishing block-diagonal part:
\begin{equation*} \label{Tj1}
T_n^{(j)}=\left( \begin{array}{cc}
0&T^{(j)}_{12}\\
T^{(j)}_{21}&0
\end{array} \right).
\end{equation*}
After replacing $Y_n=T_n\varphi_n$ in (\ref{lineardiscrete}) we get
\begin{equation}  \label{Th}
T_{n+1}h_n=\left(\sum^{\infty}_{j=-1}f_n^{(j)}\lambda^{-j}\right)T_n,
\end{equation}
where $h_n=\varphi_{n+1}\varphi^{-1}_n$.
Let us replace in (\ref{Th}) the factors by their formal expansions:
$$(E+T_{n+1}^{(1)}\lambda^{-1}+\cdots)(h_n^{(-1)}\lambda+  h_n^{(0)}+\cdots )=(f_n^{(-1)}\lambda+f_n^{(0)}+\cdots)(E+T_n^{(1)}\lambda^{-1}+\cdots) .$$
By comparing coefficients at the powers of $\lambda$ we derive a sequence of equations
\begin{eqnarray}\label{sequence1}
h_n^{(-1)}= f_n^{(-1)}, \\
 \label{sequencek}
T_{n+1}^{(k)}h_n^{(-1)}+h_n^{(k-1)}-f_n^{(-1)} T_n^{(k)}=R^k_n,\quad k\geq1.
\end{eqnarray}
Here $R^k_n$ denotes terms that have already been found in the previous steps.

To find the unknown coefficients $T_n^{(j)}$, we must solve linear equations, that look like difference equations. However due to the special form of the coefficient $f_n^{(-1)}$ these equations are linear algebraic and therefore are solved without ``integration". In other words $T_n^{(j)}$ and $h_n^{(j)}$ are local functions of the potential since depend on a finite numbers of the shifts of the functions $f_n^{(-1)}$, $f_n^{(0)}$, $f_n^{(1)}$, etc. Indeed, equation (\ref{sequencek}) obviously implies 
\begin{equation*} \label{Tj2}
\left( \begin{array}{cc}
0&D_n(T^{(k)}_{12})A_{22}\\
0&0
\end{array} \right)+\left( \begin{array}{cc}
p&0\\
0&q
\end{array} \right)-\left( \begin{array}{cc}
0&0\\
A_{22}T^{(k)}_{21}&0
\end{array} \right)=R^k_n.
\end{equation*}
Here $D_n$ is the operator shifting the argument $n$: $D_ny_n=y_{n+1}$, and $p=h^{(k-1)}_{11}$, $q=h^{(k-1)}_{22}$.
Evidently this equation is easily solved and the searched matrices $T_n^{(k)}$ and $h_n^{(k-1)}$ are uniquely found for any $k\geq 1$.

Suppose now that the system of equations
\begin{equation}  \label{Y10Y01}
Y_{n+1,m}=(f_{n,m}^{(-1)}\lambda+f_{n,m}^{(0)}+\cdots)Y_{n,m}, \quad 
Y_{n,m+1}=G_{n,m}(\lambda)Y_{n,m},
\end{equation}
where $G_{n,m}(\lambda)$ is analytic at a vicinity of $\lambda=\infty$, is the Lax pair for the nonlinear quad system 
\begin{equation*}  \label{F}
Q([u^{j}_{n,m}])=0,
\end{equation*}
i.e. $Q$ depends on the variable $u^{j}_{n,m}$ and on its shifts with respect to the variables $j,n,m$.

Assume that function $f^{(-1)}_{n,m}$ has the structure (\ref{f_01}). Then due to the reasonings above there exists a linear transformation $Y_{n,m}\longmapsto \varphi_{n,m}=T^{-1}_{n,m}Y_{n,m}$ which reduces the first equation in (\ref{Y10Y01}) to a block-diagonal form (\ref{diag})-(\ref{coeff-h}). It can be checked that this transformation brings also the second equation of (\ref{f_01}) to an equation of the same block structure 
\begin{equation*}  \label{diagS}
\varphi_{n,m+1}=S_{n,m}\varphi_{n,m},
\end{equation*}
where $S_{n,m}$ is a formal power series 
\begin{equation*}\label{seriesS}
S_{n,m}=S_{n,m}^{(0)}+S_{n,m}^{(1)}\lambda^{-1}+S_{n,m}^{(2)}\lambda^{-2}+\cdots
\end{equation*}
and
\begin{equation*} \label{coeff-S}
S_{n,m}^{(j)}=\left( \begin{array}{cc}
S^{(j)}_{11}&0\\
0&S^{(j)}_{22}
\end{array} \right).
\end{equation*}
Since the compatibility property of linear systems is preserved under change of variables, we have the relation
\begin{equation*} \label{comp_hS}
S_{n+1,m}h_{n,m}=h_{n,m+1}S_{n,m}
\end{equation*}
which implies due to the block-diagonal structure that
\begin{equation*} \label{cl_hS}
(D_n-1)\log \det (S_{ii})=(D_m-1)\log \det (h_{ii}), \, i=1,2.
\end{equation*}
By evaluating and comparing the coefficients at the powers of $\lambda$ we derive the sequence of the local conservation laws.

The Lax pairs considered in the article have also the second singular point $\lambda=0$, so we briefly discuss the Lax pair represented as
\begin{equation}  \label{lambda0}
Y_{n+1,m}=F_{n,m}Y_{n,m}, \, 
Y_{n,m+1}=(g_{n,m}^{(-1)}\lambda^{-1}+g_{n,m}^{(0)}+g_{n,m}^{(1)}\lambda+\cdots)Y_{n,m},
\end{equation}
where $F_{n,m}=F_{n,m}(\lambda)$ is analytic at a vicinity of $\lambda=0$. We request the here the  term $g_{n,m}^{(-1)}$  has the block structure (\ref{f_02}). In this case the block-diagonalization is performed in a way very similar to the one recalled above for the singularity $\lambda=\infty$.

\subsection{Quad system corresponding to $A^{(1)}_{N-1}$}

Let us apply the scheme discussed above to the Lax pair of the quad system (\ref{A1-N}). To this end we present the Lax pair (\ref{7new}), (\ref{8new}) in a matrix form
\begin{equation}\label{78new_matrix}
\Phi_{1,0}=F\Phi, \quad \Phi_{0,1}=G\Phi,
\end{equation}
where $F$ and $G$ are as follows
\begin{equation*} \label{FA1N}
F=\left(
\begin{array}{cccccc}
\frac{t^1_{0,0}t^2_{1,0}}{t^1_{1,0}t^2_{0,0}}&-1&0&\dots&0&0\\
0&\frac{t^2_{0,0}t^{3}_{1,0}}{t^2_{1,0}t^{3}_{0,0}}&-1&\dots&0&0\\
0&0&\frac{t^3_{0,0}t^{4}_{1,0}}{t^3_{1,0}t^{4}_{0,0}}&\dots&0&0\\
&&&\dots\\
-\lambda\frac{t^1_{0,0}t^2_{1,0}}{t^1_{1,0}t^2_{0,0}}&\lambda&0&\dots&0&\frac{t^{N}_{0,0}t^1_{2,0}}{t^{N}_{1,0}t^1_{1,0}}\\
\end{array}\right),\end{equation*}
\begin{equation*} \label{GA1N}
G=\left(
\begin{array}{ccccccc}
a\frac{t^1_{0,0}t^1_{1,1}}{t^1_{1,0}t^1_{0,1}}&0&0&\dots&0&b^2\lambda^{-1}\frac{t^2_{0,1}t^{N-1}_{0,0}}{t^1_{0,1}t^N_{0,0}}&b\lambda^{-1}\frac{t^2_{0,1}t^N_{0,1}}{t^1_{1,0}t^1_{0,1}}\\
\frac{bt^1_{0,0}t^3_{0,1}}{t^2_{0,0}t^2_{0,1}}&1&0&\dots&0&0&0\\
0&\frac{bt^2_{0,0}t^4_{0,1}}{t^3_{0,0}t^3_{0,1}}&1&\dots&0&0&0\\
&&\dots\\
0&0&0&\dots&0&\frac{bt^{N-1}_{0,0}t^1_{1,0}}{t^{N}_{0,0}t^{N}_{0,1}}&1\\
\end{array}\right).
\end{equation*}
First we reduce the Lax pair (\ref{78new_matrix}) to the appropriate form (\ref{Y10Y01}), (\ref{f_01}) (or to the form (\ref{lambda0}), (\ref{f_02})) by means of the transformation 
\begin{equation*}
\Phi=HY \quad (\mbox{or}\quad  \Phi=\bar HY), 
\end{equation*}
where the factor $H$ is lower (respectively $\bar H$ is upper) block-diagonal matrix
\begin{equation*} \label{H}
H=\left( \begin{array}{cc}
H_{11}&0\\
H_{21}&H_{22}
\end{array} \right),\quad \left(\mbox{or}\quad 
\bar H=\left( \begin{array}{cc}
\bar H_{11}&\bar H_{12}\\
0&\bar H_{22}
\end{array} \right)\right).
\end{equation*}
Here $H_{11}$, $H_{22}$, $\bar H_{11}$, $\bar H_{22}$ are diagonal matrices, some entries of which might depend on the spectral parameter $\xi=\lambda^{\frac{1}{N-1}}$ (or $\zeta=\lambda^{-\frac{1}{N-1}}$). Examples show that these factors are effectively found. Below, for simplicity, we illustrate all the arguments and calculations using the example of $N=3$.

{\bf Example 1.} For $N=3$ the quad system (\ref{A1-N}) reads as 
\begin{equation} \label{discA12}
\left\{ \begin{array}{c}
at^1_{0,0}t^1_{1,1}-t^1_{1,0}t^1_{0,1}=bt^2_{0,1}t^3_{0,1},\\
at^2_{0,0}t^2_{1,1}-t^2_{1,0}t^2_{0,1}=bt^1_{1,0}t^3_{0,1},\\
at^3_{0,0}t^3_{1,1}-t^3_{1,0}t^3_{0,1}=bt^1_{1,0}t^2_{1,0}.
\end{array} \right.
\end{equation}
System (\ref{discA12}) corresponds to the algebra $A^{(1)}_2$ and relates  with the Lax pair 
\begin{equation}  \label{LaxA12}
\Phi_{1,0}=F\Phi, \quad  \Phi_{0,1}=G\Phi,
\end{equation}
where 
\begin{equation*} \label{fgA12}
F=\left( \begin{array}{ccc}
\frac{t^1_{0,0}t^2_{1,0}}{t^1_{1,0}t^2_{0,0}}&-1&0\\
0&\frac{t^2_{0,0}t^3_{1,0}}{t^2_{1,0}t^3_{0,0}}&-1\\
-\frac{t^1_{0,0}t^2_{1,0}}{t^1_{1,0}t^2_{0,0}}\lambda&\lambda&\frac{at^1_{2,0}t^3_{0,0}}{t^1_{1,0}t^3_{1,0}}
\end{array} \right), \quad 
G=\left( \begin{array}{ccc}
\frac{at^1_{0,0}t^1_{1,1}}{t^1_{1,0}t^1_{0,1}}&\frac{b^2t^2_{0,0}t^2_{0,1}}{\lambda t^1_{0,1}t^3_{0,0}}&\frac{bt^2_{0,1}t^3_{0,1}}{\lambda t^1_{1,0}t^1_{0,1}}\\
\frac{bt^1_{0,0}t^3_{0,1}}{t^2_{0,0}t^2_{0,1}}& 1&0\\
0&\frac{bt^2_{0,0}t^1_{1,0}}{t^3_{0,0}t^3_{0,1}}&1
\end{array} \right).
\end{equation*}
Let us perform the formal diagonalization (really block-diagonalization) procedure to the system (\ref{LaxA12}) around the singular values $\lambda=0$ and $\lambda=\infty$ of the spectral parameter. We begin with the case $\lambda=\infty$.  By changing the variables 
\begin{equation*} \label{zamA12}
\Phi=HY, 
\end{equation*}
where
\begin{equation*} \label{HA12}
H=\left( \begin{array}{ccc}
1&0&0 \\ 
\frac{t^1_{0,0}t^2_{1,0}}{t^1_{1,0}t^2_{0,0}}&\xi^{-1}&0 \\
\frac{t^1_{0,0}t^3_{1,0}}{t^1_{1,0}t^3_{0,0}}&0&1
\end{array} \right), \quad \xi=\sqrt{\lambda},
\end{equation*}
we reduce (\ref{LaxA12}) to the form
\begin{equation}  \label{Y10Y01A12}
Y_{1,0}=fY, \quad 
Y_{0,1}=gY.
\end{equation}
Here $f=f^{(-1)}\xi+f^{(0)}+f^{(1)}\xi^{-1}$, $g=g^{(0)}+g^{(1)}\xi^{-1}+g^{(2)}\xi^{-2}+g^{(3)}\xi^{-3}$,
\begin{equation*} \label{fm1f0A12}
f^{(-1)}=\left( \begin{array}{ccc}
0&0&0\\ 0&0&-1\\ 0&1&0
\end{array} \right), \quad
f^{(0)}=\left( \begin{array}{ccc}
0&0&0\\ 
0&\frac{t^1_{1,0}t^2_{2,0}}{t^1_{2,0}t^2_{1,0}}+\frac{t^2_{0,0}t^3_{1,0}}{t^2_{1,0}t^3_{0,0}}&0\\
\frac{at^1_{0,0}t^1_{2,0}}{(t^1_{1,0})^2}&0&\frac{at^1_{2,0}t^3_{0,0}}{t^1_{1,0}t^3_{1,0}}
\end{array} \right),
\end{equation*}
\begin{equation*} \label{f1g0A12}
f^{(1)}=\left( \begin{array}{ccc}
0&-1&0\\ 0&0&0\\ 0&\frac{t^1_{1,0}t^3_{2,0}}{t^1_{2,0}t^3_{1,0}}&0
\end{array} \right), \quad
g^{(0)}=\left( \begin{array}{ccc}
\frac{at^1_{0,0}t^1_{1,1}}{t^1_{1,0}t^1_{0,1}}&0&0\\ 0&1&0\\ 0&0&1
\end{array} \right),
\end{equation*}
\begin{equation*} \label{g1A12}
g^{(1)}=\left( \begin{array}{ccc}
0&0&0\\ -\frac{abt^1_{0,0}t^2_{1,1}t^3_{1,1}}{(t^1_{1,0})^2t^1_{1,1}}&0&-\frac{bt^2_{1,1}t^3_{0,1}}{t^1_{1,0}t^1_{1,1}}\\ 0&\frac{bt^1_{1,0}t^2_{0,0}}{t^3_{0,0}t^3_{0,1}}&0
\end{array} \right),
\end{equation*}
\begin{equation*} \label{g2g3A12}
g^{(2)}=\left( \begin{array}{ccc}
\frac{abt^1_{0,0}t^2_{0,1}t^3_{1,1}}{(t^1_{1,0})^2t^1_{0,1}}&0&\frac{bt^2_{0,1}t^3_{0,1}}{t^1_{1,0}t^1_{0,1}} \\ 
0&-\frac{b^2t^2_{0,0}t^2_{1,1}}{t^3_{0,0}t^1_{1,1}}&0\\ 
-\frac{abt^1_{0,0}t^2_{0,1}(t^3_{1,1})^2}{(t^1_{1,0})^2t^1_{1,1}t^3_{0,1}}&0&-\frac{bt^2_{0,1}t^3_{1,1}}{t^1_{1,0}t^1_{1,1}}
\end{array} \right), \quad 
g^{(3)}=\left( \begin{array}{ccc}
0&\frac{b^2t^2_{0,0}t^2_{0,1}}{t^3_{0,0}t^1_{0,1}}&0 \\ 
0&0&0\\ 
0&-\frac{b^2t^2_{0,0}t^2_{0,1}t^3_{1,1}}{t^3_{0,0}t^3_{0,1}t^1_{1,1}}&0
\end{array} \right).
\end{equation*}
Obviously the leading term $f^{(-1)}$ of the potential $f$ at $\xi\mapsto\infty$ is of the necessary block-diagonal form. According to the above scheme, there is a formal series
\begin{equation*}
T=E+T^{(1)}\xi^{-1}+T^{(2)}\xi^{-2}+\cdots \label{series2_TA12}
\end{equation*}
with the coefficients having the following block structure
\begin{equation*} \label{TiA12}
T^{(i)}=\left( \begin{array}{ccc}
0&*&* \\ 
*&0&0 \\
*&0&0
\end{array} \right), \quad i\geq1,
\end{equation*}
such that replacement $Y=T\varphi$ brings the system (\ref{Y10Y01A12}) to a block-diagonal form
\begin{equation}  \label{fi10fi01A12}
\varphi_{1,0}=h\varphi, \quad 
\varphi_{0,1}=S\varphi,
\end{equation}
where the potentials $h$ and $S$ are also formal series:
\begin{eqnarray*}\label{series2_hSA12}
h=h^{(-1)}\xi + h^{(0)}+h^{(1)}\xi^{-1}+h^{(2)}\xi^{-2}+\cdots,\\
S=S^{(0)}+S^{(1)}\xi^{-1}+S^{(2)}\xi^{-2}+\cdots.
\end{eqnarray*}
Here we assume that the potential $h$ has a block diagonal form 
\begin{equation*} \label{hiA12}
h^{(i)}=\left( \begin{array}{ccc}
*&0&0 \\ 
0&*&* \\
0&*&*
\end{array} \right), i\geq -1.
\end{equation*}
Then we can check that $S^{(j)}$ for all $j\geq0$ has the same block diagonal matrix structure
\begin{equation*} \label{SiA12}
S^{(j)}=\left( \begin{array}{ccc}
*&0&0 \\ 
0&*&* \\
0&*&*
\end{array} \right).
\end{equation*}
Omitting the calculations we give several members of these series
\begin{eqnarray*}\label{TneqA12}
T=\left( \begin{array}{ccc} 1&0&0 \\ 0&1&0 \\ 0&0&1 \end{array}\right)+\left( \begin{array}{ccc} 0&0&0 \\ -\frac{at^1_{0,0}t^1_{2,0}}{(t^1_{1,0})^2}&0&0 \\ 0&0&0 \end{array}\right)\xi^{-1}+\nonumber\\
+\left( \begin{array}{ccc} 0&0&-1 \\ 0&0&0 \\ -\frac{at^1_{0,0}(t^1_{1,0}t^2_{2,0}t^3_{0,0}+t^1_{2,0}t^2_{0,0}t^3_{1,0})}{(t^1_{1,0})^2t^2_{1,0}t^3_{0,0}}&0&0 \end{array}\right)\xi^{-2}+\nonumber\\
+\left( \begin{array}{ccc} 0&-\frac{at^1_{1,0}t^3_{-1,0}}{t^1_{0,0}t^3_{0,0}}&0 \\ \frac{at^1_{0,0}t^3_{2,0}}{t^1_{1,0}t^3_{1,0}}+\frac{a^2t^1_{0,0}t^1_{2,0}t^2_{2,0}t^3_{0,0}}{(t^1_{1,0})^2t^2_{1,0}t^3_{1,0}}+\frac{a^2t^1_{0,0}(t^1_{2,0})^2t^2_{0,0}}{(t^1_{1,0})^3t^2_{1,0}}&0&0 \\ 0&0&0 \end{array}\right)\xi^{-3}+\dots,
\end{eqnarray*}
\begin{eqnarray*}\label{hneqA12}
h=\left( \begin{array}{ccc} 0&0&0 \\ 0&0&-1 \\ 0&1&0 \end{array}\right)\xi+\left( \begin{array}{ccc} 0&0&0 \\ 0&\frac{t^1_{1,0}t^2_{2,0}}{t^1_{2,0}t^2_{1,0}}+\frac{t^2_{0,0}t^3_{1,0}}{t^2_{1,0}t^3_{0,0}}&0 \\ 0&0&\frac{at^1_{2,0}t^3_{0,0}}{t^1_{1,0}t^3_{1,0}} \end{array}\right)+\left( \begin{array}{ccc} 0&0&0 \\ 0&0&0 \\ 0&\frac{t^1_{1,0}t^3_{2,0}}{t^1_{2,0}t^3_{1,0}}&0 \end{array}\right)\xi^{-1}+\nonumber\\ 
+\left( \begin{array}{ccc} \frac{at^1_{0,0}t^1_{2,0}}{(t^1_{1,0})^2}&0&0 \\ 0&0&0 \\ 0&0&-\frac{at^1_{0,0}t^1_{2,0}}{(t^1_{1,0})^2} \end{array}\right)\xi^{-2}+\left( \begin{array}{ccc} 0&0&0 \\ 0&0&0 \\ 0&-\frac{a^2t^1_{2,0}t^3_{-1,0}}{t^1_{1,0}t^3_{0,0}}&0 \end{array}\right)\xi^{-3}+\nonumber\\
+\left( \begin{array}{ccc} -\frac{at^1_{0,0}}{t^1_{1,0}}\left(\frac{t^3_{2,0}}{t^3_{1,0}}+\frac{at^1_{2,0}t^2_{2,0}t^3_{0,0}}{t^1_{1,0}t^2_{1,0}t^3_{1,0}}+\frac{a(t^1_{2,0})^2t^2_{0,0}}{(t^1_{1,0})^2t^2_{1,0}}\right)&0&0 \\ 0&0&0 \\ 0&0&h^4_{33} \end{array}\right)\xi^{-4}+\dots,
\end{eqnarray*}
where $h^4_{33}=\frac{at^1_{0,0}t^1_{2,0}}{t^1_{1,0}t^3_{0,0}}\left(\frac{t^1_{0,0}t^3_{1,0}}{(t^1_{1,0})^2}+\frac{at^2_{1,0}t^3_{-1,0}}{t^1_{1,0}t^2_{0,0}}+\frac{at^2_{-1,0}t^3_{0,0}}{t^1_{0,0}t^2_{0,0}}\right)$,
\begin{eqnarray*}\label{SneqA12}
S=\left( \begin{array}{ccc} \frac{at^1_{0,0}t^1_{1,1}}{t^1_{1,0}t^1_{0,1}}&0&0 \\ 0&1&0 \\ 0&0&1 \end{array}\right)+\left( \begin{array}{ccc} 0&0&0 \\ 0&0&-\frac{bt^2_{1,1}t^3_{0,1}}{t^1_{1,0}t^1_{1,1}} \\ 0&\frac{bt^1_{1,0}t^2_{0,0}}{t^3_{0,0}t^3_{0,1}}&0 \end{array}\right)\xi^{-1}+\nonumber\\ 
+\left( \begin{array}{ccc} \frac{abt^1_{0,0}t^2_{0,1}t^3_{1,1}}{(t^1_{1,0})^2t^1_{0,1}}&0&0 \\ 0&-\frac{b^2t^2_{0,0}t^2_{1,1}}{t^1_{1,1}t^3_{0,0}}&0 \\ 0&0&-\frac{bt^2_{0,1}t^3_{1,1}}{t^1_{1,0}t^1_{1,1}} \end{array}\right)\xi^{-2}+\left( \begin{array}{ccc} 0&0&0 \\ 0&0&\frac{abt^1_{0,0}t^2_{1,1}t^3_{1,1}}{(t^1_{1,0})^2t^1_{1,1}} \\ 0&-\frac{b^2t^2_{0,0}t^2_{0,1}t^3_{1,1}}{t^1_{1,1}t^3_{0,0}t^3_{0,1}}&0 \end{array}\right)\xi^{-3}+\nonumber\\
+\left( \begin{array}{ccc} -\frac{abt^1_{0,0}t^2_{0,1}(at^1_{2,0}t^2_{0,0}t^3_{1,1}+t^1_{1,0}t^2_{2,0}t^3_{0,1})}{(t^1_{1,0})^3t^1_{0,1}t^2_{1,0}}&0&0 \\ 0&\frac{a^2bt^3_{-1,0}t^2_{1,1}t^3_{1,1}}{t^1_{1,0}t^1_{1,1}t^3_{0,0}}&0 \\ 0&0&\frac{abt^1_{0,0}t^2_{0,1}(t^3_{1,1})^2}{(t^1_{1,0})^2t^1_{1,1}t^3_{0,1}} \end{array}\right)\xi^{-4}+\dots.
\end{eqnarray*}
The consistency condition of the system (\ref{fi10fi01A12}) can be written as $D_n(S)h=D_m(h)S$. Due to the representation 
\begin{eqnarray*}\label{hSA12}
h=\left( \begin{array}{cc} h_{11}&0 \\ 0&h_{22} \end{array}\right), \quad
S=\left( \begin{array}{cc} S_{11}&0 \\ 0&S_{22} \end{array}\right)
\end{eqnarray*}
the latter implies 
\begin{eqnarray}\label{formconslawsA12}
\left(D_n-1\right)\log\det(S_{ii})=\left(D_m-1\right)\log\det(h_{ii}), \quad i=1,2.
\end{eqnarray}
We can derive an infinite set of the local conservation laws from the equation (\ref{formconslawsA12}). Let us give some of them
\begin{enumerate}
	\item[1.] $\left(D_n-1\right)\left(-\frac{bt^2_{0,1}t^3_{1,1}}{t^1_{1,0}t^1_{1,1}}\right)=\left(D_m-1\right)\left(\frac{t^1_{1,0}t^3_{2,0}}{t^1_{2,0}t^3_{1,0}}+\frac{at^2_{2,0}t^3_{0,0}}{t^2_{1,0}t^3_{1,0}}+\frac{at^1_{2,0}t^2_{0,0}}{t^1_{1,0}t^2_{1,0}}\right)$,
	\item[2.] $\left(D_n-1\right)\left(-\frac{abt^1_{2,0}t^2_{0,0}t^2_{0,1}t^3_{1,1}}{(t^1_{1,0})^2t^2_{1,0}t^1_{1,1}}-\frac{bt^2_{2,0}t^2_{0,1}t^3_{0,1}}{t^1_{1,0}t^2_{1,0}t^1_{1,1}}-\frac{1}{2}\frac{b^2(t^2_{0,1}t^3_{1,1})^2}{(t^1_{1,0}t^1_{1,1})^2}\right)=$
	\item[] $\left(D_m-1\right)\left(\frac{at^1_{1,0}t^2_{3,0}}{t^1_{2,0}t^2_{2,0}}+\frac{at^1_{1,0}t^1_{3,0}t^2_{1,0}t^3_{2,0}}{(t^1_{2,0})^2t^2_{2,0}t^3_{1,0}}+\frac{a^2t^1_{3,0}t^3_{0,0}}{t^1_{2,0}t^3_{1,0}}+\frac{1}{2}\left(\frac{t^1_{1,0}t^3_{2,0}}{t^1_{2,0}t^3_{1,0}}+\frac{at^2_{2,0}t^3_{0,0}}{t^2_{1,0}t^3_{1,0}}+\frac{at^1_{2,0}t^2_{0,0}}{t^1_{1,0}t^2_{1,0}}\right)^2\right)$,
	\item[3.] $\left(D_n-1\right)\left(\frac{b^3t^1_{0,0}t^3_{1,1}}{t^1_{1,1}t^3_{0,0}}-\frac{a^2bt^2_{1,1}t^3_{-1,0}t^3_{1,1}}{t^1_{1,0}t^1_{1,1}t^3_{0,0}}-\frac{bt^1_{0,0}t^2_{0,1}t^3_{1,0}t^3_{1,1}}{(t^1_{1,0})^2t^1_{1,1}t^3_{0,0}}+\frac{1}{2}\frac{(bt^2_{0,1}t^3_{1,1})^2}{(t^1_{1,0}t^1_{1,1})^2}\right)=$
	\item[] $\left(D_m-1\right)\left(\frac{at^1_{0,0}t^2_{2,0}}{t^1_{1,0}t^2_{1,0}}+\frac{a^2t^1_{2,0}t^3_{-1,0}}{t^1_{1,0}t^3_{0,0}}+\frac{at^1_{0,0}t^1_{2,0}t^2_{0,0}t^3_{1,0}}{(t^1_{1,0})^2t^2_{1,0}t^3_{0,0}}+\frac{1}{2}\left(\frac{t^1_{1,0}t^3_{2,0}}{t^1_{2,0}t^3_{1,0}}+\frac{at^2_{2,0}t^3_{0,0}}{t^2_{1,0}t^3_{1,0}}+\frac{at^1_{2,0}t^2_{0,0}}{t^1_{1,0}t^2_{1,0}}\right)^2\right)$.
\end{enumerate}

In a similar way we study the system (\ref{LaxA12}) around  the point $\lambda=0$. In this case we use the change of the variables  $\Phi=\bar{H}Y$
\begin{equation*} \label{HmA12}
\bar{H}=\left( \begin{array}{ccc}
1&0&\frac{t^3_{0,0}t^2_{0,1}}{b^2t^1_{0,0}t^1_{1,0}}\\ 0&\zeta^{-1}&-\frac{t^3_{0,0}t^3_{0,1}}{bt^2_{0,0}t^1_{1,0}}\\ 0&0&1
\end{array} \right), \quad \zeta=\frac{1}{\sqrt{\lambda}}\rightarrow\infty
\end{equation*}
and arrive at the system 
\begin{equation*}  \label{Y10Y01mA12}
Y_{1,0}=\bar{f}Y, \quad 
Y_{0,1}=\bar{g}Y
\end{equation*}
with potentials
\begin{equation*}
\bar{g}=\bar{g}^{(-1)}\zeta+\bar{g}^{(0)}+\bar{g}^{(1)}\zeta^{-1}\quad \mbox{and} \quad \bar{f}=\bar{f}^{(0)}+\bar{f}^{(1)}\zeta^{-1}+\bar{f}^{(2)}\zeta^{-2}+\bar{f}^{(3)}\zeta^{-3},
\end{equation*}
where
\begin{equation*} \label{Gm1G0A12}
\bar{g}^{(-1)}=\left( \begin{array}{ccc}
0&\frac{b^2t^2_{0,0}t^2_{0,1}}{t^1_{0,1}t^3_{0,0}}&0 \\ \frac{bt^1_{0,0}t^3_{0,1}}{t^2_{0,0}t^2_{0,1}}&0&0 \\ 0&0&0
\end{array} \right), \quad
\bar{g}^{(0)}=\left( \begin{array}{ccc}
\frac{at^1_{0,0}t^1_{1,1}}{t^1_{1,0}t^1_{0,1}}&0&\frac{at^2_{0,1}t^3_{0,0}t^1_{1,1}}{b^2(t^1_{1,0})^2t^1_{0,1}} \\ 0&1+\frac{t^1_{1,0}t^2_{0,0}t^3_{02}}{t^2_{0,1}t^3_{0,0}t^1_{1,1}}&0 \\ 0&0&0
\end{array} \right),
\end{equation*}
\begin{equation*} \label{G1f0A12}
\bar{g}^{(1)}=\left( \begin{array}{ccc}
0&-\frac{t^1_{1,0}t^2_{0,0}t^2_{0,2}}{bt^1_{0,1}t^1_{1,1}t^3_{0,0}}&0 \\ 0&0&0 \\ 0&\frac{bt^1_{1,0}t^2_{0,0}}{t^3_{0,0}t^3_{0,1}}&0
\end{array} \right), \quad
\bar{f}^{(0)}=\left( \begin{array}{ccc}
\frac{t^1_{0,0}t^2_{1,0}}{t^1_{1,0}t^2_{0,0}}&0&0 \\ 0&\frac{t^2_{0,0}t^3_{1,0}}{t^2_{1,0}t^3_{0,0}}&0 \\ 0&0&\frac{at^3_{0,0}t^1_{2,0}}{t^1_{1,0}t^3_{1,0}}
\end{array} \right),
\end{equation*}
\begin{equation*} \label{f1A12}
\bar{f}^{(1)}=\left( \begin{array}{ccc}
0&-1&0 \\ -\frac{t^1_{0,0}t^3_{1,0}t^3_{1,1}}{bt^1_{1,0}t^1_{2,0}t^2_{0,0}}&0&-\frac{at^2_{1,1}t^3_{0,0}t^3_{1,0}t^3_{1,1}}{b^3(t^1_{1,0})^2t^1_{2,0}t^2_{1,0}} \\ 0&0&0
\end{array} \right),
\end{equation*}
\begin{equation*} \label{f2f3A12}
\bar{f}^{(2)}=\left( \begin{array}{ccc}
\frac{t^1_{0,0}t^2_{1,0}t^2_{1,1}t^3_{1,0}}{b^2(t^1_{1,0})^2t^1_{2,0}t^2_{0,0}}&0&\frac{a(t^2_{1,1})^2t^3_{0,0}t^3_{1,0}}{b^4(t^1_{1,0})^3t^1_{2,0}} \\ 0&\frac{t^3_{1,0}t^3_{1,1}}{bt^1_{2,0}t^2_{1,0}}&0 \\ -\frac{t^1_{0,0}t^2_{1,0}}{t^1_{1,0}t^2_{0,0}}&0&-\frac{at^2_{1,1}t^3_{0,0}}{b^2(t^1_{1,0})^2}
\end{array} \right), \quad
\bar{f}^{(3)}=\left( \begin{array}{ccc}
0&-\frac{t^3_{1,0}t^2_{1,1}}{b^2t^1_{1,0}t^1_{2,0}}&0 \\ 0&0&0 \\ 0&1&0
\end{array} \right).
\end{equation*}
Now the singular point is $\zeta=\infty$ and the coefficient $\bar g^{(-1)}$ at $\zeta$ is of the necessary block-diagonal form. Therefore we can find the formal series $\bar{T}$ and $\bar{h}$ from the equation 
\begin{equation*}\label{formThm}
\bar{T}_{0,1}\bar{h}=(\bar{g}^{(-1)}\zeta+\bar{g}^{(0)}+\bar{g}^{(1)}\zeta^{-1})\bar{T}.
\end{equation*}
Then knowing $\bar{T}$ we find the series $\bar{S}$ from 
\begin{equation*}\label{formSm}
\bar{T}_{1,0}\bar{S}=\bar{f}\bar{T}.
\end{equation*}
As a result we obtain
\begin{eqnarray*}\label{TmeqA12}
\bar{T}=\left( \begin{array}{ccc} 1&0&0 \\ 0&1&0 \\ 0&0&1 \end{array}\right)+\left( \begin{array}{ccc} 0&0&0 \\ 0&0&-\frac{at^1_{1,1}(t^3_{0,0})^2}{b^4(t^1_{1,0})^2t^2_{0,0}} \\ 0&0&0 \end{array}\right)\zeta^{-1}+\nonumber\\
+\left( \begin{array}{ccc} 0&0&\frac{at^1_{1,1}t^2_{0,1}(t^3_{0,0})^2}{b^5t^1_{0,0}(t^1_{1,0})^2t^3_{0,1}}+\frac{at^2_{0,0}t^3_{0,0}t^3_{0,2}}{b^5t^1_{0,0}t^1_{1,0}t^3_{0,1}} \\ 0&0&0 \\ \frac{at^1_{0,-1}t^1_{1,0}}{bt^2_{0,0}t^3_{0,0}}-1&0&0 \end{array}\right)\zeta^{-2}+\nonumber\\
+\left( \begin{array}{ccc} 0&0&0 \\ 0&0&-\frac{at^2_{0,2}(t^3_{0,0})^2}{b^7t^1_{1,0}t^2_{0,0}t^2_{0,1}}-\frac{a^2(t^1_{1,1})^2(t^3_{0,0})^3}{b^7(t^1_{1,0})^3t^3_{0,1}t^2_{0,0}}-\frac{a^2t^1_{1,1}(t^3_{0,0})^2t^3_{0,2}}{b^7(t^1_{1,0})^2t^3_{0,1}t^2_{0,1}} \\ 0&-\frac{at^1_{1,0}t^2_{0,-1}}{(bt^3_{0,0})^2}&0 \end{array}\right)\zeta^{-3}+\dots,
\end{eqnarray*}
\begin{eqnarray*}\label{hmeqA12}
\bar{h}=\left( \begin{array}{ccc} 0&\frac{b^2t^2_{0,0}t^2_{0,1}}{t^1_{0,1}t^3_{0,0}}&0 \\ \frac{bt^1_{0,0}t^3_{0,1}}{t^2_{0,0}t^2_{0,1}}&0&0 \\ 0&0&0 \end{array}\right)\zeta+\left( \begin{array}{ccc} \frac{at^1_{0,0}t^1_{1,1}}{t^1_{1,0}t^1_{0,1}}&0&0 \\ 0&1+\frac{t^1_{1,0}t^2_{0,0}t^3_{0,2}}{t^1_{1,1}t^2_{0,1}t^3_{0,0}}&0 \\ 0&0&0 \end{array}\right)+\nonumber\\
+\left( \begin{array}{ccc} 0&-\frac{t^1_{1,0}t^2_{0,0}t^2_{0,2}}{bt^1_{0,1}t^1_{1,1}t^3_{0,0}}&0 \\ 0&0&0 \\ 0&0&0 \end{array}\right)\zeta^{-1} 
+\left( \begin{array}{ccc} \frac{a^2t^1_{0,-1}t^1_{1,1}t^2_{0,1}}{b^3t^1_{1,0}t^1_{0,1}t^2_{0,0}}-\frac{at^1_{1,1}t^2_{0,1}t^3_{0,0}}{b^2(t^1_{1,0})^2t^1_{0,1}}&0&0 \\ 0&0&0 \\ 0&0&-\frac{at^1_{1,1}t^3_{0,0}}{b^3t^1_{1,0}t^3_{0,1}} \end{array}\right)\zeta^{-2}+\nonumber\\
+\left( \begin{array}{ccc} 0&-\frac{a^2t^1_{1,1}t^2_{0,-1}t^2_{0,1}}{b^4t^1_{1,0}t^1_{0,1}t^3_{0,0}}&0 \\ 0&0&0 \\ 0&0&0 \end{array}\right)\zeta^{-3}+\dots,
\end{eqnarray*}
\begin{eqnarray*}\label{SmeqA12}
\bar{S}=\left( \begin{array}{ccc} \frac{t^1_{0,0}t^2_{1,0}}{t^1_{1,0}t^2_{0,0}}&0&0 \\ 0&\frac{t^2_{0,0}t^3_{1,0}}{t^2_{1,0}t^3_{0,0}}&0 \\ 0&0&\frac{at^1_{2,0}t^3_{0,0}}{t^1_{1,0}t^3_{1,0}} \end{array}\right)+\left( \begin{array}{ccc} 0&-1&0 \\ -\frac{t^1_{0,0}t^3_{1,0}t^3_{1,1}}{bt^1_{1,0}t^1_{2,0}t^2_{0,0}}&0&0 \\ 0&0&0 \end{array}\right)\zeta^{-1}+\nonumber\\ 
+\left( \begin{array}{ccc} \frac{t^1_{0,0}t^2_{1,0}t^2_{1,1}t^3_{1,0}}{b^2(t^1_{1,0})^2t^1_{2,0}t^2_{0,0}}&0&0 \\ 0&\frac{t^3_{1,0}t^3_{1,1}}{bt^1_{2,0}t^2_{1,0}}&0 \\ 0&0&-\frac{at^2_{1,1}t^3_{0,0}}{(bt^1_{1,0})^2} \end{array}\right)\zeta^{-2}+\nonumber\\
+\left( \begin{array}{ccc} 0&-\frac{t^2_{1,1}t^3_{1,0}}{b^2t^1_{1,0}t^1_{2,0}}&0 \\ -\frac{at^2_{1,1}t^3_{1,0}t^3_{1,1}(at^1_{0,-1}t^1_{1,0}-bt^2_{0,0}t^3_{0,0})}{b^4(t^1_{1,0})^2t^1_{2,0}t^2_{0,0}t^2_{1,0}}&0&0 \\ 0&0&0 \end{array}\right)\zeta^{-3}+\dots.
\end{eqnarray*}
Passing to the blocks
\begin{eqnarray*}
\bar{h}=\left( \begin{array}{cc} \bar{h}_{11}&0 \\ 0&\bar{h}_{22} \end{array}\right), \quad
\bar{S}=\left( \begin{array}{cc} \bar{S}_{11}&0 \\ 0&\bar{S}_{22} \end{array}\right)
\end{eqnarray*}
we can write the relations 
\begin{eqnarray*}\label{formconslawsA12m}
\left(D_n-1\right)\log\det(\bar{S}_{ii})=\left(D_m-1\right)\log\det(\bar{h}_{ii}), \quad i=1,2,
\end{eqnarray*}
from which we can derive the local conservation laws:
\begin{enumerate}
	\item[1.] $\left(D_n-1\right)\left(\frac{t^1_{1,0}t^2_{0,2}}{bt^1_{1,1}t^2_{0,1}}+\frac{at^2_{0,0}t^3_{0,2}}{bt^2_{0,1}t^3_{0,1}}+\frac{at^1_{1,1}t^3_{0,0}}{bt^1_{1,0}t^3_{0,1}}\right)=\left(D_m-1\right)\left(-\frac{t^2_{1,1}t^3_{1,0}}{t^1_{1,0}t^1_{2,0}}\right)$,
	\item[2.] $\left(D_n-1\right)\left(\frac{1}{2}\frac{(t^1_{1,0}t^2_{0,2})^2}{(t^1_{1,1}t^2_{0,1})^2}-\frac{at^1_{0,0}t^1_{1,0}(t^2_{0,2})^2}{t^1_{0,1}t^1_{1,1}(t^2_{0,1})^2}-\frac{a^2t^1_{0,0}t^2_{0,0}t^2_{0,2}t^3_{0,2}}{t^1_{0,1}(t^2_{0,1})^2t^3_{0,1}}-\frac{t^1_{1,0}t^1_{0,2}t^2_{0,2}t^3_{0,1}}{t^1_{0,1}t^1_{1,1}t^2_{0,1}t^3_{0,2}}-\frac{1}{2}\frac{(at^2_{0,0}t^3_{0,2})^2}{(t^2_{0,1}t^3_{0,1})^2}-\frac{a^2t^1_{1,1}t^2_{0,0}t^3_{0,0}t^3_{0,2}}{t^1_{1,0}t^2_{0,1}(t^3_{0,1})^2}\right.$\\
	\item[] $\left.-\frac{1}{2}\frac{(at^1_{1,1}t^3_{0,0})^2}{(t^1_{1,0}t^3_{0,1})^2}-\frac{at^1_{0,2}t^2_{0,0}}{t^1_{0,1}t^2_{0,1}}-\frac{at^1_{1,0}t^3_{0,3}}{t^1_{1,1}t^3_{0,2}}-\frac{at^2_{0,2}t^3_{0,0}}{t^2_{0,1}t^3_{0,1}}\right)=\left(D_m-1\right)\left(\frac{bt^2_{1,0}t^3_{1,0}t^3_{0,2}}{t^1_{1,0}t^1_{2,0}t^3_{0,1}}+\frac{abt^1_{1,1}t^2_{1,1}t^3_{0,0}t^3_{1,0}}{(t^1_{1,0})^2t^1_{2,0}t^3_{0,1}}+\frac{1}{2}\frac{(bt^2_{1,1}t^3_{1,0})^2}{(t^1_{1,0}t^1_{2,0})^2}\right)$,
	\item[3.] $\left(D_n-1\right)\left[\frac{at^1_{0,-1}t^1_{1,0}-bt^2_{0,0}t^3_{0,0}}{t^1_{0,0}t^1_{1,0}t^3_{0,1}}\left(\frac{at^3_{0,2}t^1_{1,0}t^2_{0,0}+at^1_{1,1}t^2_{0,1}t^3_{0,0}}{t^1_{1,0}t^2_{0,0}}\right)+\frac{a^2t^1_{1,1}t^2_{0,-1}}{t^1_{1,0}t^2_{0,0}}+\frac{1}{2}\frac{(t^1_{1,0}t^2_{0,2})^2}{(t^1_{1,1}t^2_{0,1})^2}+\frac{at^1_{1,0}t^2_{0,0}t^2_{0,2}t^3_{0,2}}{t^1_{1,1}(t^2_{0,1})^2t^3_{0,1}}+\frac{at^2_{0,2}t^3_{0,0}}{t^2_{0,1}t^3_{0,1}}\right.$
	\item[] $\left.+\frac{1}{2}\frac{a^2(t^1_{1,0}t^2_{0,0}t^3_{0,2}+t^1_{1,1}t^2_{0,1}t^3_{0,0})^2}{(t^1_{1,0}t^2_{0,1}t^3_{0,1})^2}\right]=$
	\item[] $\left(D_m-1\right)\left(\frac{ab^3t^1_{0,-1}t^1_{1,0}t^2_{1,1}}{t^1_{0,0}t^1_{2,0}t^2_{0,0}}-\frac{a^2bt^2_{0,-1}t^2_{1,1}t^3_{1,1}}{t^1_{1,0}t^1_{2,0}t^2_{0,0}}-\frac{b^4t^2_{1,1}t^3_{0,0}}{t^1_{0,0}t^1_{2,0}}-\frac{abt^1_{0,-1}t^2_{0,1}t^2_{1,1}t^3_{1,0}}{t^1_{0,0}t^1_{1,0}t^1_{2,0}t^2_{0,0}}+\frac{b^2t^2_{0,1}t^2_{1,1}t^3_{0,0}t^3_{1,0}}{t^1_{0,0}(t^1_{1,0})^2t^1_{2,0}}+\frac{1}{2}\frac{(bt^3_{1,0}t^2_{1,1})^2}{(t^1_{1,0}t^1_{2,0})^2}\right)$.
\end{enumerate}

\section{Higher symmetries}

Quad system (\ref{A1-N}) possess higher symmetries. However, presented in the variables $t^j_{n,m}$ the symmetries have non-localities. They become local in new variables introduced by potentiation or, more precisely, by setting $r^j_{0,0}=\frac{t^j_{1,0}}{t^j_{0,0}}$. Actually in terms of these variables quad system (\ref{A1-N}) turns into 
\begin{equation} \label{eqA1N}
\left. \begin{array}{l}
ar^1_{1,1}=r^1_{1,0}+r^2_{0,1}r^N_{0,1}\left(\frac{a}{r^1_{0,0}}-\frac{1}{r^1_{0,1}}\right),\\
ar^j_{1,1}=r^j_{1,0}+r^{j-1}_{1,0}r^{j+1}_{0,1}\left(\frac{a}{r^j_{0,0}}-\frac{1}{r^j_{0,1}}\right), \quad 2\leq j\leq N-1,\\
ar^N_{1,1}=r^N_{1,0}+r^1_{1,0}r^{N-1}_{1,0}\left(\frac{a}{r^N_{0,0}}-\frac{1}{r^N_{0,1}}\right).
\end{array} \right.
\end{equation}
More precisely in what follows we present higher symmetries to this quad system. 
We note that obtained system  (\ref{eqA1N}) does not already contain the parameter b.
Everywhere below we write $r^{j}$ instead of $r^{j}_{0,0}$.

The Lax pair of the system (\ref{eqA1N}) is written as
\begin{eqnarray}
\Phi_{1,0}=F\Phi,\label{7new_laxr}\\
\Phi_{0,1}=G\Phi,\label{8new_laxr}
\end{eqnarray}
where $F$ and $G$ are matrices
\begin{equation*} \label{FA1Nr}
F=\left(
\begin{array}{cccccc}
\frac{r^2}{r^1}&-1&0&\dots&0&0\\
0&\frac{r^3}{r^2}&-1&\dots&0&0\\
0&0&\frac{r^4}{r^3}&\dots&0&0\\
&&&\dots\\
-\lambda\frac{r^2}{r^1}&\lambda&0&\dots&0&\frac{ar^1_{1,0}}{r^{N}}\\
\end{array}\right),\end{equation*}
\begin{equation*} \label{GA1Nr}
G=\left(
\begin{array}{ccccccc}
\frac{ar^1_{0,1}}{r^1}&0&0&\dots&0&\lambda^{-1}\frac{(ar^1_{0,1}-r^1)(ar^N_{0,1}-r^N)}{r^1r^{N-1}}&\lambda^{-1}\frac{ar^1_{0,1}-r^1}{r^1}\\
\frac{ar^2_{0,1}-r^2}{r^1}&1&0&\dots&0&0&0\\
0&\frac{ar^3_{0,1}-r^3}{r^2}&1&\dots&0&0&0\\
&&\dots\\
0&0&0&\dots&0&\frac{ar^N_{0,1}-r^N}{r^{N-1}}&1\\
\end{array}\right).
\end{equation*}

{\bf Example 2.} Consider the system (\ref{eqA1N}) for $N=2$. For the sake of simplicity here we use notations $u:=r^1$, $v:=r^2$
\begin{equation} \label{eqA11}
\left. \begin{array}{l}
au_{1,1}=u_{1,0}+v^2_{0,1}\left(\frac{a}{u}-\frac{1}{u_{0,1}}\right),\\
av_{1,1}=v_{1,0}+u^2_{1,0}\left(\frac{a}{v}-\frac{1}{v_{0,1}}\right).
\end{array} \right.
\end{equation} 
The simplest higher symmetry of (\ref{eqA11}) in the direction of $n$ is given by
\begin{equation} \label{symteqA11}
\left. \begin{array}{l}
u_{t}=v+\frac{au^2}{v_{-1,0}},\\
v_{t}=au_{1,0}+\frac{v^2}{u}.
\end{array} \right.
\end{equation}
The symmetry admits Lax pair $\Phi_{1,0}=F\Phi$, $\Phi_t=A\Phi$ where 
\begin{eqnarray*}\label{AfA11}
F=\left( \begin{array}{cc} \frac{v}{u}&-1 \\ -\lambda\frac{v}{u} & \lambda+\frac{au_{1,0}}{v} \end{array} \right),\,
A=\left( \begin{array}{cc} \frac{au}{v_{-1,0}}+\lambda & 1 \\ \frac{v}{u}\lambda & \frac{v}{u} \end{array}\right).
\end{eqnarray*}

For the particular case $a=1$, quad system (\ref{eqA11}) and also its symmetry (\ref{symteqA11}) have already been found in \cite{GHY}. We also indicate the symmetry in the direction of $m$ \cite{HabKhPavl}
\begin{equation} \label{symtaueqA11}
\left. \begin{array}{l}
u_{\tau}=\frac{u^2}{av_{0,1}-v},\\
v_{\tau}=\frac{uu_{0,-1}}{au-u_{0,-1}}
\end{array} \right.
\end{equation}
with the Lax pair $\Phi_{0,1}=G\Phi$, $\Phi_{\tau}=B\Phi$ where
\begin{eqnarray*}\label{gBA11}
G=\left( \begin{array}{cc} \frac{au_{0,1}}{u}+\frac{(a u_{0,1}-u)(av_{0,1}-v)}{\lambda u^2}& \frac{au_{0,1}-u}{\lambda u} \\ \frac{av_{0,1}-v}{u} & 1 \end{array} \right),\,
B=\left( \begin{array}{cc} \frac{u}{av_{0,1}-v}& \frac{u}{\lambda(av_{0,1}-v)} \\ \frac{u_{0,-1}}{au-u_{0,-1}}&-\frac{1}{\lambda}\end{array}\right).
\end{eqnarray*}

For the special choice $a=1$ of the parameter  the quad system (\ref{eqA11}) and its symmetries take the most simple form. Namely the coupled lattice (\ref{symteqA11}) in that case is converted to a scalar lattice
\begin{equation*}
R_{n,m,t}=R_{n+1,m}+\frac{R^2_{n,m}}{R_{n-1,m}}
\end{equation*}
where $R_{2n,m}=u_{n,m}$ and $R_{2n+1,m}=v_{n,m}$. Hence the desire to transform the discrete system itself to a scalar equation. However, the symmetry (\ref{symtaueqA11}) in the other direction  is reduced to a scalar form 
\begin{equation*}
P_{n,m,\tau}=\frac{1}{P_{n,m-1}-P_{n,m+1}}
\end{equation*}
under the different transformation $P_{n,2m}=\frac{1}{u_{n,m}}$, $P_{n,2m-1}=v_{n,m}$. Therefore the system of equations  (\ref{eqA11}) is not reduced to a scalar autonomous  equation  \cite{Yamilov2019}.  If $a\neq 1$ then none of the symmetries is reduced to an autonomous scalar lattice.

{\bf Example 3.}  Now we represent the higher symmetries to the system (\ref{eqA1N}) for $N=3$.  Here we use notations $u:=r^1$, $v:=r^2$ and $w:=r^3$
\begin{equation} \label{eqA12}
\left. \begin{array}{l}
au_{1,1}=u_{1,0}+v_{0,1}w_{0,1}\left(\frac{a}{u}-\frac{1}{u_{0,1}}\right),\\
av_{1,1}=v_{1,0}+u_{1,0}w_{0,1}\left(\frac{a}{v}-\frac{1}{v_{0,1}}\right),\\
aw_{1,1}=w_{1,0}+u_{1,0}v_{1,0}\left(\frac{a}{w}-\frac{1}{w_{0,1}}\right).\\
\end{array} \right.
\end{equation}
The Lax pair to the system takes the form 
\begin{equation}  \label{LaxA12uvw}
\Phi_{1,0}=F\Phi, \quad  \Phi_{0,1}=G\Phi,
\end{equation}
where 
\begin{equation*} \label{fgA12uvw}
F=\left( \begin{array}{ccc}
\frac{v}{u}&-1&0\\
0&\frac{w}{v}&-1\\
-\frac{v}{u}\lambda&\lambda&\frac{au_{1,0}}{w}
\end{array} \right), \quad 
G=\left( \begin{array}{ccc}
\frac{au_{0,1}}{u}&\frac{(au_{0,1}-u)(aw_{0,1}-w)}{\lambda uv}&\frac{au_{0,1}-u}{\lambda u}\\
\frac{av_{0,1}-v}{u}& 1&0\\
0&\frac{aw_{0,1}-w}{v}&1
\end{array} \right).
\end{equation*}

The simplest higher symmetry of the system (\ref{eqA12}) is given by
\begin{equation} \label{symteqA12}
\left. \begin{array}{l}
u_{t}=w+\frac{auv}{w_{-1,0}}+\frac{au^2}{v_{-1,0}},\\
v_{t}=au_{1,0}+\frac{vw}{u}+\frac{av^2}{w_{-1,0}},\\
w_{t}=av_{1,0}+\frac{au_{1,0}w}{v}+\frac{w^2}{u}.
\end{array} \right.
\end{equation}
The linear system of the form  
\begin{equation*}\label{lax-symteqA12}
\Phi_{1,0}=F\Phi,\quad \Phi_{t}=A\Phi
\end{equation*}
is a Lax pair for (\ref{symteqA12}), where $F$ is given in (\ref{LaxA12uvw}) and $A$ is as
\begin{equation*} \label{AA12uvw}
A=\left( \begin{array}{ccc}
\lambda+\frac{au}{v_{-1,0}}&\frac{au}{w_{-1,0}}&1\\
\frac{v}{u}\lambda&\frac{av}{w_{-1,0}}&\frac{v}{u}\\
\frac{w}{u}\lambda&0&\frac{w}{u}
\end{array} \right).
\end{equation*}
The quad system (\ref{eqA12}) admits also the classical symmetries
$$u_{t}=u,\quad v_{t}=v,\quad w_{t}=w$$
and
$$u_{t}=(-1)^nu,\quad v_{t}=(-1)^{n+1}v,\quad w_{t}=(-1)^nw.$$

In the direction of $m$ the system  has a symmetry of the form
\begin{equation*} \label{symtaueqA12}
\left. \begin{array}{l}
u_{\tau}=\frac{u^2v}{(av_{0,1}-v)(aw_{0,1}-w)},\\
v_{\tau}=\frac{uvu_{0,-1}}{(aw_{0,1}-w)(au-u_{0,-1})},\\
w_{\tau}=\frac{uu_{0,-1}v_{0,-1}}{(au-u_{0,-1})(av-v_{0,-1})}.
\end{array} \right.
\end{equation*}
The symmetry is related to the Lax pair $\Phi_{0,1}=G\Phi$, $\Phi_{\tau}=B\Phi$ with $G$ and $B$ defined by (\ref{LaxA12uvw}) and, respectively, by
\begin{eqnarray*}\label{BA12}
B=\left( \begin{array}{ccc} \frac{uv}{(av_{0,1}-v)(aw_{0,1}-w)}& 0 & \frac{uv}{(av_{0,1}-v)(aw_{0,1}-w)}\lambda^{-1} \\ \frac{vu_{0,-1}}{(aw_{0,1}-w)(au-u_{0,-1})}&0 & -\frac{v}{aw_{0,1}-w}\lambda^{-1}\\ -\frac{u_{0,-1}}{au-u_{0,-1}} & \frac{auu_{0,-1}}{(au-u_{0,-1})(av-v_{0,-1})} & \lambda^{-1} \end{array}\right).
\end{eqnarray*}

{\bf Example 4.} The next example concerns the algebra $A^{(1)}_3$. In this case the quad system (\ref{eqA1N}) reads as 
\begin{equation} \label{eqA13}
\left. \begin{array}{l}
ar^1_{1,1}=r^1_{1,0}+r^2_{0,1}r^4_{0,1}\left(\frac{a}{r^1}-\frac{1}{r^1_{0,1}}\right),\\
ar^2_{1,1}=r^2_{1,0}+r^1_{1,0}r^3_{0,1}\left(\frac{a}{r^2}-\frac{1}{r^2_{0,1}}\right),\\
ar^3_{1,1}=r^3_{1,0}+r^2_{1,0}r^4_{0,1}\left(\frac{a}{r^3}-\frac{1}{r^3_{0,1}}\right),\\
ar^4_{1,1}=r^4_{1,0}+r^1_{1,0}r^3_{1,0}\left(\frac{a}{r^4}-\frac{1}{r^4_{0,1}}\right).
\end{array} \right.
\end{equation}
The Lax pair for (\ref{eqA13}) is given by a system of the form 
\begin{equation*}  \label{LaxA13}
\Phi_{1,0}=F\Phi, \quad  \Phi_{0,1}=G\Phi,
\end{equation*}
where the potentials are
\begin{equation*}\label{feqA13}
F=\left( \begin{array}{cccc} \frac{r^2}{r^1}&-1&0&0 \\ 0&\frac{r^3}{r^2}&-1&0 \\ 0&0&\frac{r^4}{r^3}&-1 \\ -\lambda\frac{r^2}{r^1}&\lambda&0&\frac{ar^1_{1,0}}{r^4} \end{array}\right),
\end{equation*}
\begin{equation*}\label{geqA13}
G=\left( \begin{array}{cccc} \frac{ar^1_{0,1}}{r^1}&0&\frac{(ar^1_{0,1}-r^1)(ar^4_{0,1}-r^4)}{r^1r^3}\lambda^{-1}&\frac{ar^1_{0,1}-r^1}{r^1}\lambda^{-1} \\ \frac{ar^2_{0,1}-r^2}{r^1}&1&0&0 \\ 0&\frac{ar^3_{0,1}-r^3}{r^2}&1&0 \\ 0&0&\frac{ar^4_{0,1}-r^4}{r^3}&1 \end{array}\right).
\end{equation*}
The following multifield lattices are symmetries for the quad system (\ref{eqA13})
\begin{equation*} \label{symteqA13}
\left. \begin{array}{l}
r^1_{t}=r^4+\frac{a(r^1)^2}{r^2_{-1,0}}+\frac{ar^1r^2}{r^3_{-1,0}}+\frac{ar^1r^3}{r^4_{-1,0}},\\
r^2_{t}=ar^1_{1,0}+\frac{a(r^2)^2}{r^3_{-1,0}}+\frac{ar^2r^3}{r^4_{-1,0}}+\frac{r^2r^4}{r^1},\\
r^3_{t}=ar^2_{1,0}+\frac{a(r^3)^2}{r^4_{-1,0}}+\frac{r^3r^4}{r^1}+\frac{ar^3r^1_{1,0}}{r^2},\\
r^4_{t}=ar^3_{1,0}+\frac{(r^4)^2}{r^1}+\frac{ar^4r^1_{1,0}}{r^2}+\frac{ar^4r^2_{1,0}}{r^3}
\end{array} \right.
\end{equation*}
and
\begin{equation*} \label{symtaueqA13}
\left. \begin{array}{l}
r^1_{\tau}=\frac{(r^1)^2r^2r^3}{(ar^2_{0,1}-r^2)(ar^3_{0,1}-r^3)(ar^4_{0,1}-r^4)},\\
r^2_{\tau}=\frac{r^1r^2r^3r^1_{0,-1}}{(ar^1-r^1_{0,-1})(ar^3_{0,1}-r^3)(ar^4_{0,1}-r^4)},\\
r^3_{\tau}=\frac{r^1r^3r^1_{0,-1}r^2_{0,-1}}{(ar^1-r^1_{0,-1})(ar^2-r^2_{0,-1})(ar^4_{0,1}-r^4)},\\
r^4_{\tau}=\frac{r^1r^1_{0,-1}r^2_{0,-1}r^3_{0,-1}}{(ar^1-r^1_{0,-1})(ar^2-r^2_{0,-1})(ar^3-r^3_{0,-1})}.
\end{array} \right.
\end{equation*}
The Lax pairs for these lattices $\Phi_{1,0}=F\Phi$, $\Phi_t=A\Phi$ and $\Phi_{0,1}=G\Phi$, $\Phi_{\tau}=B\Phi$ are given by
\begin{eqnarray*}\label{AA13}
A=\left( \begin{array}{cccc} \frac{ar^1}{r^2_{-1,0}}+\lambda & \frac{ar^1}{r^3_{-1,0}} & \frac{ar^1}{r^4_{-1,0}} & 1\\ \frac{r^2}{r^1}\lambda & \frac{ar^2}{r^3_{-1,0}} & \frac{ar^2}{r^4_{-1,0}} & \frac{r^2}{r^1} \\ \frac{r^3}{r^1}\lambda & 0 & \frac{ar^3}{r^4_{-1,0}} & \frac{r^3}{r^1} \\ \frac{r^4}{r^1}\lambda& 0& 0& \frac{r^4}{r^1}\end{array}\right), 
\end{eqnarray*}
\begin{eqnarray*}\label{BA13}
B=\left( \begin{array}{cccc} \frac{-r^1r^2r^3}{P^{(3)}(r^2,r^3,r^4)}& 0 & 0 & \frac{-r^1r^2r^3\lambda^{-1}}{P^{(3)}(r^2,r^3,r^4)}\\ \frac{-r^1_{0,-1}r^2r^3}{P^{(3)}(r^1_{0,-1},r^3,r^4)}&0 & 0 & \frac{r^2r^3\lambda^{-1}}{P^{(2)}(r^3,r^4)}\\ \frac{r^1_{0,-1}r^3}{P^{(2)}(r^1_{0,-1},r^4)} & \frac{-ar^1r^1_{0,-1}r^3}{P^{(3)}(r^1_{0,-1},r^2_{0,-1},r^4)} & 0 & \frac{-r^3\lambda^{-1}}{ar^4_{0,1}-r^4}\\ \frac{-r^1_{0,-1}}{ar^1-r^1_{0,-1}} & \frac{ar^1r^1_{0,-1}}{P^{(2)}(r^1_{0,-1},r^2_{0,-1})} & \frac{-ar^1r^1_{0,-1}r^2_{0,-1}}{P^{(3)}(r^1_{0,-1},r^2_{0,-1},r^3_{0,-1})} & \lambda^{-1}\end{array}\right),
\end{eqnarray*}
where $P^{(3)}(h,k,l)=(ah_{0,1}-h)(ak_{0,1}-k)(al_{0,1}-l),P^{(2)}(h,k)=(ah_{0,1}-h)(ak_{0,1}-k).$

{\bf Example 5.} Finally, we concentrate on the general case of the system (\ref{eqA1N}). Symmetries of the system are given as
\begin{equation*} \label{symteqA1Nm1}
\left. \begin{array}{l}
r^1_{t}=r^N+\frac{a(r^1)^2}{r^2_{-1,0}}+\frac{ar^1r^2}{r^3_{-1,0}}+\frac{ar^1r^3}{r^4_{-1,0}}+\dots+\frac{ar^1r^{N-1}}{r^N_{-1,0}},\\
r^2_{t}=ar^1_{1,0}+\frac{a(r^2)^2}{r^3_{-1,0}}+\frac{ar^2r^3}{r^4_{-1,0}}+\frac{ar^2r^4}{r^5_{-1,0}}+\dots+\frac{ar^2r^{N-1}}{r^N_{-1,0}}+\frac{r^2r^N}{r^1},\\
r^3_{t}=ar^2_{1,0}+\frac{a(r^3)^2}{r^4_{-1,0}}+\frac{ar^3r^4}{r^5_{-1,0}}+\dots+\frac{ar^3r^{N-1}}{r^N_{-1,0}}+\frac{r^3r^N}{r^1}+\frac{ar^3r^1_{1,0}}{r^2},\\
\dots,\\
r^{N-1}_{t}=ar^{N-2}_{1,0}+\frac{a(r^{N-1})^2}{r^N_{-1,0}}+\frac{r^{N-1}r^N}{r^1}+\frac{ar^{N-1}r^1_{1,0}}{r^2}+\dots+\frac{ar^{N-1}r^{N-3}_{1,0}}{r^{N-2}},\\
r^{N}_{t}=ar^{N-1}_{1,0}+\frac{(r^{N})^2}{r^1}+\frac{ar^{N}r^1_{1,0}}{r^2}+\frac{ar^{N}r^2_{1,0}}{r^3}+\dots+\frac{ar^{N}r^{N-2}_{1,0}}{r^{N-1}}
\end{array} \right.
\end{equation*}
and
\begin{equation*} \label{symtaueqA1Nm1}
\left. \begin{array}{l}
r^1_{\tau}=\frac{(r^1)^2r^2r^3\dots r^{N-1}}{(ar^2_{0,1}-r^2)(ar^3_{0,1}-r^3)(ar^4_{0,1}-r^4)\dots(ar^N_{0,1}-r^N)},\\
r^2_{\tau}=\frac{r^1r^2r^3\dots r^{N-1}r^1_{0,-1}}{(ar^1-r^1_{0,-1})(ar^3_{0,1}-r^3)(ar^4_{0,1}-r^4)\dots(ar^N_{0,1}-r^N)},\\
r^3_{\tau}=\frac{r^1r^3\dots r^{N-1}r^1_{0,-1}r^2_{0,-1}}{(ar^1-r^1_{0,-1})(ar^2-r^2_{0,-1})(ar^4_{0,1}-r^4)\dots(ar^N_{0,1}-r^N)},\\
\dots,\\
r^{N-1}_{\tau}=\frac{r^1r^{N-1}r^1_{0,-1}r^2_{0,-1}r^3_{0,-1}\dots r^{N-2}_{0,-1}}{(ar^1-r^1_{0,-1})(ar^2-r^2_{0,-1})\dots(ar^{N-2}-r^{N-2}_{0,-1})(ar^N_{0,1}-r^N)},\\
r^{N}_{\tau}=\frac{r^1r^1_{0,-1}r^2_{0,-1}r^3_{0,-1}\dots r^{N-1}_{0,-1}}{(ar^1-r^1_{0,-1})(ar^2-r^2_{0,-1})(ar^3-r^3_{0,-1})\dots(ar^{N-1}-r^{N-1}_{0,-1})}.
\end{array} \right.
\end{equation*}

System (\ref{eqA1N}) looks very similar to a system of quad equations studied earlier in \cite{Doliwa} (see (2.1) in \cite{Doliwa}). Let us compare in detail the two systems. We present that from \cite{Doliwa} in the form 
\begin{equation} \label{systemsDoliwa}
\left. \begin{array}{l}
p^j_{n,m+1}=p^j_{n+1,m}+\frac{p^j_{n,m+1}p^j_{n+1,m}}{p^j_{n+1,m+1}p^{j+1}_{n,m}}\left(p^{j+1}_{n,m+1}-p^{j+1}_{n+1,m}\right), \quad j=1,\dots,N-1,\\
p^N_{n,m+1}=p^N_{n+1,m}+\frac{p^N_{n,m+1}p^N_{n+1,m}}{p^N_{n+1,m+1}p^{1}_{n,m}}\left(p^{1}_{n,m+1}-p^{1}_{n+1,m}\right).
\end{array} \right.
\end{equation}
Evidently (\ref{systemsDoliwa}) is a periodic reduction 
\begin{equation} \label{periodicDoliwa}
p^{j+N}_{n,m}=p^{j}_{n,m}
\end{equation}
of the Hirota-Miwa equation written as 
\begin{equation} \label{HMDoliwa}
\left. \begin{array}{l}
p^j_{n,m+1}=p^j_{n+1,m}+\frac{p^j_{n,m+1}p^j_{n+1,m}}{p^j_{n+1,m+1}p^{j+1}_{n,m}}\left(p^{j+1}_{n,m+1}-p^{j+1}_{n+1,m}\right), \, -\infty<j<\infty.
\end{array} \right.
\end{equation}
In turn, system (\ref{eqA1N}) can be obtained from another form of the Hirota-Miwa equation 
\begin{equation} \label{HMr}
ar^j_{n+1,m+1}=r^j_{n+1,m}+\frac{r^{j-1}_{n+1,m}r^{j+1}_{n,m+1}}{r^j_{n,m}r^j_{n,m+1}}\left(ar^j_{n,m+1}-r^j_{n,m}\right), \, -\infty<j<\infty
\end{equation}
by imposing a restriction
\begin{equation} \label{constraintA1N}
r^{j+N}_{n,m}=r^{j}_{n+1,m-1}.
\end{equation}
It can easily be checked that in the case $a=1$ equations (\ref{HMDoliwa}) and (\ref{HMr}) are related by
\begin{equation} \label{A1NDoliwa}
p^{j}_{n,m}=r^{n+m-1}_{j,m}.
\end{equation}
For $a\neq 1$ the equations are essentially different. Direct computation convinces that triple of the equations (\ref{periodicDoliwa}), (\ref{constraintA1N}), (\ref{A1NDoliwa}) is not consistent, therefore

{\bf Proposition 3.} The systems (\ref{eqA1N}) and (\ref{systemsDoliwa}) are not related by a point transformation.

\section{Evaluation of the recursion operators}

In this section we discuss the recursion operators for the quad systems associated with $A^{(1)}_{N-1}$. Recursion operators are closely related with higher symmetries, local conservation laws and multi-hamiltonian structures. There are several methods for constructing the recursion operators, based on the Lax representation \cite{Gurses}, \cite{Mikh}, Hamiltonian operators \cite{Fuchssteiner}, \cite{Maltsev} and generalized invariant manifolds \cite{HKhJPA2018}. In our opinion the most convenient of them is one using the Lax pair, moreover in some cases it is reasonable to derive the bihamiltonian structure from the recursion operator. 

Since the quad system admits two hierarchies of symmetries, it admits two recursion operators as well, corresponding to the variables $n$ and $m$. Below we concentrate on that related to $n$. To study the problem we use the ideas of \cite{Gurses}, \cite{Mikh}.

It can be shown that the procedure of the formal diagonalization allows us to construct a formal series (see \cite{HabYang})
\begin{equation*}\label{seriesA}
A=A^{(0)}+A^{(1)}\lambda^{-1}+A^{(2)}\lambda^{-2}+\dots
\end{equation*}
commuting with the operator $L=D_n^{-1}F$ such that 
\begin{equation*}\label{commutLA}
\left[L,A\right]:=LA-AL=0,
\end{equation*}
where $F=F^{(0)}+F^{(1)}\lambda$ is the potential of the Lax equation (\ref{7new_laxr}) and the expression $D_n^{-1}F$ means a composition of the backward shift operator $D_n^{-1}$ and the operation of multiplication by $F$.

The series $A$ provides an effective tool for constructing the higher symmetries of the quad system (\ref{eqA1N}). To explain the method we consider the formal series $B$ obtained from $A$ by multiplying by $\lambda^k$, where $k$ is a positive integer:
\begin{equation}\label{seriesB}
B:=\lambda^{k}A=B^{(k)}\lambda^{k}+B^{(k-1)}\lambda^{k-1}+\dots.
\end{equation}
Then we take a polynomial part of the series (\ref{seriesB})
\begin{equation*}\label{seriesBplus}
B_+=B^{(k)}\lambda^{k}+B^{(k-1)}\lambda^{k-1}+\dots+B^{(1)}\lambda+B^{(0)}_+.
\end{equation*}
Here the last summand is chosen in a nontrivial way. Its upper diagonal part coincides with that of the matrix $B^{(0)}$. The other entries vanish except one located at the left upper corner, denote it by $x$. The crucial point is that the consistency condition of the linear equation 
\begin{equation}\label{ytauBplus}
\frac{d}{d\tau}y=B_+y
\end{equation}
with (\ref{7new_laxr}) gives exactly $N+1$ equations which allow us to determine the unknown $x$ and to derive a symmetry of the quad system (\ref{eqA1N}) with time $\tau$.

In addition to (\ref{seriesB}) we take one more series $C:=\lambda^{k+1}A$, such that
\begin{equation}\label{seriesC}
C=C^{(k+1)}\lambda^{k+1}+C^{(k)}\lambda^{k}+C^{(k-1)}\lambda^{k-1}+\dots.
\end{equation}
By applying the algorithm used above to (\ref{seriesC}) we get the polynomial 
\begin{equation*}\label{seriesCplus}
C_+=C^{(k+1)}\lambda^{k+1}+C^{(k)}\lambda^{k}+\dots+C^{(1)}\lambda+C^{(0)}_+
\end{equation*}
which provides the time part of the Lax pair
\begin{equation}\label{ytau1Cplus}
\frac{d}{d\tau_1}y=C_+y.
\end{equation}
Our goal now is to find the relation between two symmetries defined by the polynomials $B_+$, $C_+$. Evidently we have 
\begin{equation}\label{CminusB}
C-\lambda B=0.
\end{equation}
We replace in (\ref{CminusB}) the summands $B$ and $C$ with their representations (\ref{seriesB}) and (\ref{seriesC}). As a result we get 
\begin{eqnarray}\label{CminusBext}
C^{(k+1)}\lambda^{k+1}+\dots+C^{(1)}\lambda+C^{(0)}_{+}+C^{(0)}_{-}+C^{(-1)}\lambda^{-1}+\dots-\nonumber\\
-\lambda\left(B^{(k)}\lambda^{k}+\dots+B^{(1)}\lambda+B^{(0)}_{+}+B^{(0)}_{-}+B^{(-1)}\lambda^{-1}+\dots\right)=0,
\end{eqnarray}
where $B^{(0)}_{-}:=B^{(0)}-B^{(0)}_{+}$, $C^{(0)}_{-}:=C^{(0)}-C^{(0)}_{+}$.

Let us rewrite (\ref{CminusBext}) as follows 
\begin{eqnarray*}\label{Rn}
R_N:=C^{(k+1)}\lambda^{k+1}+\dots+C^{(1)}\lambda+C^{(0)}_{+}-\lambda\left(B^{(k)}\lambda^{k}+\dots+B^{(1)}\lambda+B^{(0)}_{+}\right)=\nonumber\\
\qquad -C^{(0)}_{-}-C^{(-1)}\lambda^{-1}-\dots+\lambda\left(B^{(0)}_{-}+B^{(-1)}\lambda^{-1}+\dots\right).
\end{eqnarray*}
It is easily checked that $R_N$ is a linear function of $\lambda$:
\begin{eqnarray}\label{Rn2}
R_N=B^{(0)}_{-}\lambda+C^{(0)}_{+}:=r\lambda+s.
\end{eqnarray}
The consistency conditions of the equations (\ref{7new_laxr}), (\ref{ytauBplus}) and (\ref{7new_laxr}), (\ref{ytau1Cplus}) can be written in the form 
\begin{eqnarray*}\label{Ltau1Ltau}
L_{\tau_1}=\left[L,C_+\right], \quad L_{\tau}=\left[L,B_+\right].
\end{eqnarray*}
Therefore we obtain
\begin{eqnarray*}\label{Ltau1minusLtau}
L_{\tau_1}-\lambda L_{\tau}=\left[L,C_+-\lambda B_+\right]=\left[L,R_N\right].
\end{eqnarray*}
Since $L=(\alpha+\beta\lambda)D_n^{-1}$, where $\alpha=D_n^{-1}\left(F^{(0)}\right)$ and $\beta=D_n^{-1}\left(F^{(1)}\right)$ (recall that $F=F^{(0)}+F^{(1)}\lambda$) then due to (\ref{Rn2}) we have an equation 
\begin{eqnarray*}\label{Ltau1minusLtau_ext}
\left(\alpha_{\tau_1}+\left(\beta_{\tau_1}-\alpha_{\tau}\right)\lambda-\beta_{\tau}\lambda^2\right)D_n^{-1}=\left[(\alpha+\beta\lambda)D_n^{-1},r\lambda+s\right].
\end{eqnarray*}
which produces a system of the equations 
\begin{eqnarray}\label{eq_abrs}
\alpha_{\tau_1}=\alpha s_{-1}-s\alpha,\nonumber\\
\beta_{\tau_1}-\alpha_{\tau}=\alpha r_{-1}+\beta s_{-1}-r\alpha-s\beta,\\
\beta_{\tau}=r\beta-\beta r_{-1}\nonumber
\end{eqnarray}
for determining unknown coefficients $r$ and $s$ and a relation between the symmetries $\left\{r^j_\tau\right\}^N_{j=1}$ and $\left\{r^j_{\tau_1}\right\}^N_{j=1}$, which should allow us to derive the recursion operator.

\subsection{Examples}

Let us construct the recursion operator in the direction of $n$ for a particular case of the system (\ref{eqA1N}) for $N=2$. In an explicit form the system is presented in (\ref{eqA11}). In this case, functions $\alpha$ and $\beta$ used above have the form
\begin{equation} \label{abA11}
\alpha=\left( \begin{array}{cc} \frac{u_{-1,0}}{v_{-1,0}}&-1 \\ 0 & \frac{v}{u_{-1,0}} \end{array} \right),\,
\beta=\left( \begin{array}{cc} 0&0 \\ -\frac{u_{-1,0}}{v_{-1,0}} & 1 \end{array} \right).
\end{equation}

We look for the coefficients $r$ and $s$ in the form
\begin{equation} \label{rsA11}
r=\left( \begin{array}{cc} r^{(11)} & 0 \\ r^{(21)} & 0 \end{array} \right),\,
s=\left( \begin{array}{cc} s^{(11)} & s^{(12)} \\ 0 & s^{(22)} \end{array} \right).
\end{equation}

Let's rewrite system (\ref{eq_abrs}) taking into account (\ref{abA11}) and (\ref{rsA11}):
\begin{enumerate}
	\item[1)] \quad $\left(\frac{u_{-1,0}}{v_{-1,0}}\right)_{\tau_1}-\frac{u_{-1,0}}{v_{-1,0}}\left(s^{(11)}_{-1,0}-s^{(11)}\right)=0$,
	\item[2)] \quad $s^{(22)}_{-1,0}-s^{(11)}+\frac{v}{u_{-1,0}}s^{(12)}-\frac{u_{-1,0}}{v_{-1,0}}s^{(12)}_{-1,0}=0$,
	\item[3)] \quad $\left(\frac{v}{u_{-1,0}}\right)_{\tau_1}-\frac{v}{u_{-1,0}}\left(s^{(22)}_{-1,0}-s^{(22)}\right)=0$,
	\item[4)] \quad $\left(\frac{u_{-1,0}}{v_{-1,0}}\right)_{\tau}-r^{(21)}_{-1,0}+\frac{u_{-1,0}}{v_{-1,0}}r^{(11)}_{-1,0}-\frac{u_{-1,0}}{v_{-1,0}}r^{(11)}+\frac{u_{-1,0}}{v_{-1,0}}s^{(12)}=0$,
	\item[5)] \quad $s^{(12)}-r^{(11)}=0$,
	\item[6)] \quad $\left(\frac{u_{-1,0}}{v_{-1,0}}\right)_{\tau_1}-\frac{u_{-1,0}}{v_{-1,0}}s^{(11)}_{-1,0}+\frac{u_{-1,0}}{v_{-1,0}}s^{(22)}+\frac{v}{u_{-1,0}}r^{(21)}_{-1,0}-\frac{u_{-1,0}}{v_{-1,0}}r^{(21)}=0$,
	\item[7)] \quad $\left(\frac{v}{u_{-1,0}}\right)_{\tau}-s^{(22)}+s^{(22)}_{-1,0}+r^{(21)}-\frac{u_{-1,0}}{v_{-1,0}}s^{(12)}_{-1,0}=0$,
	\item[8)] \quad $\left(\frac{u_{-1,0}}{v_{-1,0}}\right)_{\tau}-r^{(21)}_{-1,0}+\frac{u_{-1,0}}{v_{-1,0}}r^{(11)}_{-1,0}=0$.
\end{enumerate}

Thus, to find the necessary coefficients $r^{(11)}$, $r^{(21)}$, $s^{(11)}$, $s^{(12)}$ and $s^{(22)}$, we obtained an overdetermined system of equations. Using equation $5)$, we exclude  the function $s^{(12)}$ by taking $s^{(12)}=r^{(11)}$. Combining equations $1)$, $2)$, $4)$, $6)$ and $7)$ we find the function $r^{(11)}$ 
\begin{equation}\label{r11}
r^{(11)}=\frac{v_{\tau}}{v}.
\end{equation}
From equation $4)$, by virtue of (\ref{r11}), we find
\begin{equation*}\label{r21}
r^{(21)}=\frac{u_{\tau}}{v}.
\end{equation*}
We find the remaining two coefficients $s^{(11)}$ and $s^{(22)}$. To do this, use the equations $1)$, $2)$ and $6)$. From which we obtain
\begin{equation}\label{s22}
s^{(22)}=\frac{u_{\tau}}{v}+\frac{v_{\tau}}{u_{-1,0}}+\left(D_n-1\right)^{-1}\left[\left(\frac{1}{v}-\frac{v_{1,0}}{u^2}\right)u_{\tau}+\left(\frac{1}{u_{-1,0}}-\frac{u}{v^2}\right)v_{\tau}\right],
\end{equation}
\begin{equation}\label{s11}
s^{(11)}=\frac{v_{\tau}}{u_{-1,0}}+\frac{v(u_{-1,0})_{\tau}}{u_{-1,0}^2}+\left(D_n-1\right)^{-1}\left[\left(\frac{1}{v}-\frac{v_{1,0}}{u^2}\right)u_{\tau}+\left(\frac{1}{u_{-1,0}}-\frac{u}{v^2}\right)v_{\tau}\right].
\end{equation}
Let's write two more equations connecting $u_{\tau_1}$ and $u_{\tau}$, $v_{\tau_1}$ and $v_{\tau}$. Using equations $1)$ and $3)$ we obtain
\begin{equation}\label{uv}
u_{\tau_1}=-u\left(s^{(11)}_{1,0}+s^{(22)}\right), \quad v_{\tau_1}=-v\left(s^{(11)}+s^{(22)}\right).
\end{equation}
We substitute (\ref{s22}) and (\ref{s11}) in (\ref{uv}) and get
\begin{equation*}\label{uvR}
\left( \begin{array}{c} u \\ v \end{array} \right)_{\tau_1}=R\left( \begin{array}{c} u \\ v \end{array} \right)_{\tau},
\end{equation*}
where
\begin{eqnarray}\label{RA11}
&&R=\left( \begin{array}{cc} 0 & 1 \\ 0 & 0 \end{array} \right)D_n+\left( \begin{array}{cc} 2\frac{u}{v} & 2\frac{u}{u_{-1,0}}-\frac{u^2}{v^2} \\ 1 & 2\frac{v}{u_{-1,0}} \end{array} \right)+\left( \begin{array}{cc} 0 & 0 \\ \frac{v^2}{u^2_{-1,0}} & 0 \end{array} \right)D_n^{-1}\nonumber\\
&& \qquad \qquad-2\left( \begin{array}{c} u \\ v \end{array} \right)\left(D_n-1\right)^{-1}\left(\frac{v_{1,0}}{u^2}-\frac{1}{v}, \frac{u}{v^2}-\frac{1}{u_{-1,0}}\right).
\end{eqnarray}
Applying the operator (\ref{RA11}) to the classical symmetry $u_\tau=u$, $v_\tau=v$ of the system (\ref{eqA11}) we obtain the higher symmetry (\ref{symteqA11}). Operator (\ref{RA11}) has been found in our article \cite{HabKh2017}.

As a next example, we consider the system (\ref{eqA1N}) with $N=3$ (see (\ref{eqA12})). For this system, we construct the recursion operator using the method presented above. Omitting the calculations, we give only the final answer
\begin{eqnarray}\label{RA12}
R=R^{(0)}+S\left[\left(D_n-1\right)^{-1}R^{(1)}+\left(D_n+1\right)^{-1}R^{(2)}\right.\\
\left.+\left(D_n+1\right)^{-1}A^{(1)}\left(D_n-1\right)^{-1}R^{(3)}+\left(D_n+1\right)^{-1}A^{(2)}\left(D_n+1\right)^{-1}R^{(4)}\right], \nonumber
\end{eqnarray}
where
\begin{eqnarray*}\label{R0A12}
&&R^{(0)}=\left( \begin{array}{ccc} 0 & 1 & \frac{u}{v} \\ 0 & 0 & 1 \\ 0 & 0 & 0 \end{array} \right)D_n+\left( \begin{array}{ccc} 2\frac{u}{w}+\frac{w_{1,0}}{v} & \frac{u}{u_{-1,0}}-\frac{uw_{1,0}}{v^2} & 2\frac{u}{v_{-1,0}}-\frac{u^2}{w^2}+\frac{w_{1,0}}{u_{-1,0}}+\frac{uv}{u_{-1,0}w}\\  \frac{v}{w} & 2\frac{v}{u_{-1,0}}+\frac{u}{w}  & \frac{v}{v_{-1,0}} -\frac{uv}{w^2} \\1 &\frac{w}{u_{-1,0}}  & 2\frac{w}{v_{-1,0}}+\frac{v}{u_{-1,0}}  \end{array} \right)\nonumber\\
&&+\left( \begin{array}{ccc} 0 & 0 & 0\\ \frac{v^2}{u^2_{-1,0}} & 0 & 0 \\ \frac{vw}{u^2_{-1,0}} & \frac{w^2}{v^2_{-1,0}} & 0 \end{array} \right)D_n^{-1},\nonumber
\end{eqnarray*}

\begin{eqnarray*}\label{SA12}
S=\left( \begin{array}{ccc} u & 0 & 0 \\ 0 & v & 0 \\ 0 & 0 & w \end{array} \right),\\
R^{(1)}=-\frac{3}{2}\left( \begin{array}{ccc} \frac{v_{1,0}}{u^2}-\frac{1}{w} & 0 & 0 \\ 0 & \frac{w_{1,0}}{v^2}-\frac{1}{u_{-1,0}} & 0 \\ 0 & 0 & \frac{u}{w^2}-\frac{1}{v_{-1,0}} \end{array} \right),\\
R^{(2)}=\left( \begin{array}{ccc} \frac{v_{1,0}}{2u^2}-\frac{1}{2w}-\frac{w_{1,0}\alpha}{u^2}& \frac{\alpha}{w} & \frac{u_{1,0}}{u_{-1,0}v_{1,0}} +\frac{u}{w^2}-\frac{v\alpha}{w^2}-\frac{\beta}{u_{-1,0}} \\ \frac{\beta}{v} & \frac{w_{1,0}}{2v^2}-\frac{u\beta}{v^2}-\frac{1}{2u_{-1,0}} & \frac{\beta}{u_{-1,0}}+\frac{u}{w^2}+\frac{1}{v_{-1,0}} \\ 0& \frac{\gamma}{w}-\frac{1}{u_{-1,0}}-\frac{w_{1,0}}{v^2} & \frac{u}{2w^2}-\frac{1}{2v_{-1,0}}-\frac{v\gamma}{w^2}\end{array} \right),\\
R^{(3)}=\left( \begin{array}{ccc} \frac{w_{1,0}}{u^2}-\frac{1}{v} & 0 & 0 \\ 0 & \frac{u}{v^2}-\frac{1}{w} & 0 \\ 0 & 0 & \frac{v}{w^2}-\frac{1}{u_{-1,0}} \end{array} \right),\\
R^{(4)}=\left( \begin{array}{ccc} \frac{w_{1,0}}{u^2}+\frac{1}{v} & 0 & 0 \\ 0 & -\frac{u}{v^2}-\frac{1}{w} & 0 \\ 0 & 0 & \frac{v}{w^2}+\frac{1}{u_{-1,0}} \end{array} \right),\\
A^{(1)}=-\frac{1}{2}\left( \begin{array}{ccc} \frac{w_{2,0}}{u_{1,0}}+\frac{v_{1,0}}{w_{1,0}}+\frac{u}{v} & 0 & 0 \\ 0 & \frac{w_{1,0}}{u}+\frac{v}{w}+\frac{u_{1,0}}{v_{1,0}} & 0 \\ 0 & 0 & \frac{w}{u_{-1,0}}+\frac{v_{1,0}}{w_{1,0}}+\frac{u}{v} \end{array} \right),\\
A^{(2)}=\frac{1}{2}\left( \begin{array}{ccc} \frac{w_{2,0}}{u_{1,0}}+\frac{v_{1,0}}{w_{1,0}}+\frac{u}{v} & 0 & 0 \\ 0 & -\frac{w_{1,0}}{u}-\frac{v}{w}-\frac{u_{1,0}}{v_{1,0}} & 0 \\ 0 & 0 & \frac{w}{u_{-1,0}}+\frac{v_{1,0}}{w_{1,0}}+\frac{u}{v} \end{array} \right),
\end{eqnarray*}
where
$\alpha=\frac{w_{2,0}}{u_{1,0}}+\frac{v_{1,0}}{w_{1,0}}+\frac{u}{v}$, $\beta=\frac{w_{1,0}}{u}+\frac{v}{w}+\frac{u_{1,0}}{v_{1,0}}$, $\gamma=\frac{w}{u_{-1,0}}+\frac{v_{1,0}}{w_{1,0}}+\frac{u}{v}$.

Usually, recursion operators for integrable systems are pseudodifferential (or pseudo-difference for the case of lattices) operators with so-called weak nonlocalities \cite{Maltsev}. However, among lattices there are several representatives which have recursion operators with a more complex nonlocalities. This case is illustrated, for example, in \cite{Fuchssteiner} by the Narita-Itoh-Bogoyavlensky equations. Obviously, the recursion operator (\ref{RA12}) also belongs to this class, since it contains strongly non-local terms.

\section*{Conclusions}  

We conjecture that the discrete exponential type system on a quad graph (\ref{generalghy}) is an integrable discretization of the Drinfeld-Sokolov hierarchy \cite{Drinfeld}. Evidently (\ref{generalghy}) goes to (\ref{hypPDE}) in the continuum limit for appropriate choice of the parameters $a$ and $b$. As for the proof of integrability of (\ref{generalghy}), this is much more complicated problem. So far it was proved in \cite{Smirnov} that (\ref{generalghy}) is integrable in the sense of Darboux for the simple Lie algebras $A_N$, $B_N$. For the system (\ref{generalghy}) corresponding to the affine algebras $D^{(2)}_N$ and $A^{(1)}_1$ the Lax pairs were found in \cite{GHY}. For the cases $A^{(1)}_1$ and $A^{(2)}_2$ the higher symmetries have been constructed \cite{GHY}. For the cases $C_2$, $G_2$, $D_3$ the Darboux integrability was proved \cite{GHY}.

In the present article we studied the quad system corresponding to $A^{(1)}_{N-1}$. We derived for it the Lax representation, allowing to describe the local conservation laws. We constructed the higher symmetries by using this Lax pair. We derived the recursion operator for $N=3$, which turned out to be rather complicated.

\subsection*{Acknowledgements}

The authors gratefully acknowledge financial support from a Russian Science Foundation grant (project 15-11-20007).

\end{document}